\documentclass[11pt]{article}
\usepackage{fancyhdr,psfig,lastpage}
\usepackage{graphicx}
\usepackage{amssymb}
\usepackage{pstricks}
\usepackage{pst-node,pst-plot,pst-coil}
\usepackage{amsmath}
\usepackage{latexsym}
\usepackage{url}
\usepackage[sort&compress,numbers]{natbib}
\begin{document}

\vspace*{0.8cm}
\begin{center}
  {\Large \bf Evolution of Cooperation \\[2mm] 
in a Spatial  Prisoner's Dilemma }\\[10mm]
 
  {\large Frank Schweitzer$^{1,2}$, Laxmidhar Behera$^{1,3}$, Heinz M\"{u}hlenbein$^{1}$}\\[4mm]

\begin{quote}
\begin{itemize}
\item[$^{1}$] \emph{Fraunhofer Institute for Autonomous Intelligent
    Systems, \\ Schloss Birlinghoven, 53754 Sankt Augustin, Germany}
\item[$^{2}$] \emph{Institute of Physics, Humboldt University, 
    Invalidenstra{\ss}e 110, \\ 10115 Berlin, Germany,}
    \url{schweitzer@physik.hu-berlin.de} 
\item[$^{3}$] \emph{Department of Electrical Engineering, 
Indian Institute of Technology \\ Kanpur 208 016, India,} 
\url{lbehera@iitk.ac.in}
\end{itemize}
\end{quote}
\end{center}


\begin{abstract}
  We investigate the spatial distribution and the global frequency of
  agents who can either cooperate or defect. The agent interaction is
  described by a deterministic, non-iterated prisoner's dilemma game,
  further each agent only locally interacts with his neighbors. Based on
  a detailed analysis of the local payoff structures we derive critical
  conditions for the invasion or the spatial coexistence of cooperators
  and defectors.  These results are concluded in a phase diagram that
  allows to identify five regimes, each characterized by a distinct
  spatiotemporal dynamics and a corresponding final spatial structure. In
  addition to the complete invasion of defectors, we find coexistence
  regimes with either a majority of cooperators in large spatial domains,
  or a minority of cooperators organized in small non-stationary domains
  or in small clusters. The analysis further allowed a verification of
  computer simulation results by Nowak and May (1993). Eventually, we
  present simulation results of a true 5-person game on a lattice. This
  modification leads to non-uniform spatial interactions that may even
  enhance the effect of cooperation. 
\end{abstract}

\newcommand{\mean}[1]{\left\langle #1 \right\rangle}
\newcommand{\abs}[1]{\left| #1 \right|}
\newcommand{\la}{\langle}
\newcommand{\ra}{\rangle}
\newcommand{\RA}{\Rightarrow}
\newcommand{\tet}{\vartheta}
\newcommand{\eps}{\varepsilon}
\newcommand{\bbox}[1]{\mbox{\boldmath $#1$}}
\newcommand{\ul}[1]{\underline{#1}}
\newcommand{\ol}[1]{\overline{#1}}
\newcommand{\non}{\nonumber \\}
\newcommand{\no}{\nonumber}
\newcommand{\eqn}[1]{eq. (\ref{#1})}
\newcommand{\Eqn}[1]{Eq. (\ref{#1})}
\newcommand{\eqs}[2]{eqs. (\ref{#1}), (\ref{#2})}
\newcommand{\pics}[2]{Figs. \ref{#1}, \ref{#2}}
\newcommand{\pic}[1]{Fig. \ref{#1}}
\newcommand{\sect}[1]{Sect. \ref{#1}}
\newcommand{\name}[1]{{\rm #1}}
\newcommand{\vol}[1]{{\bf #1}}
\newcommand{\et}{{\it et al.}}
\newcommand{\D}{\displaystyle}
\newcommand{\T}{\textstyle}
\newcommand{\SC}{\scriptstyle}
\newcommand{\SSC}{\scriptscriptstyle}
\renewcommand{\textfraction}{0.05}
\renewcommand{\topfraction}{0.95}
\renewcommand{\bottomfraction}{0.95}
\renewcommand{\floatpagefraction}{0.95}
\section{Introduction}
\label{1}

The evolution of cooperation has been extensively studied in a
biological, social and ecological context \citep{Olson:65, Hardin:68,
  Trivers:72, Glance:93, Glance:94, Akiyama:95, Axelrod:99, Doebeli:97,
  Dugatkin:89, Lindgren:94, Darwen:02, fs-ed-02, fs-physa-02}.  In order
to obtain a general theory for this, in particular the so called
\emph{Prisoner's Dilemma} (PD) (see also \sect{2.2}) - an evolutionary
game introduced by \citet{Rapoport:65} (see also \citep{Rapoport:96}) --
has been widely investigated \citep{Boyd:87, Boyd:89, Farreli:89,
  Fogel:95, Kirchkamp:00, Lima:89, Lorberbaum:94, Michael:96,
  Muhlenbein:91, Nowak:94, Szabo:98, Grim:97}. Based on PD
investigations, in his seminal work \citet{Axelrod:84} has shown that
cooperation can emerge as a norm in a society comprised of individuals
with selfish motives -- for a review of this development in the last
twenty years see e.g.  \citep{Hoffmann:00}.

The Prisoner's Dilemma is based on the precondition that it pays off to
be \emph{not} cooperative, i.e. to defect in a cooperative environment,
this way taking a ``free ride'' at the costs of those how are
cooperating, e.g. for a commong good. This type of problem has been also
recognized as the ``tragedy of the commons'' \citep{Hardin:68}.  Many
investigations of the Prisoner's Dilemma consider a so-called
\emph{iterated game} \citep{Akiyama:95, Fogel:95, Lima:89, Darwen:02,
  Kraines:93, Oliphant:98, Matsushima:98, behera:02}, where the players
interact consecutively a given number of times, $n_{g} \geq 2$.  It makes
sense only if the players can remember the previous choices of their
opponents, i.e. if they have a memory of $n_{m}\leq n_{g}-1$ steps. Then,
they may be able to develop different \emph{strategies} -- such as the
famous ``tit for tat'' \citep{Axelrod:81, Axelrod:84} -- based on their
past experiences with their opponents.

In this paper, we are only interested in a non-iterated PD game,
$n_{g}=1$, which is also called a ``one-shot'' game. In this case the
players - or the \emph{agents} as we call them here - do not develop
strategies, they can rather choose between two different actions, to
\emph{cooperate} (C) and to \emph{defect} (D).  It can be shown (see also
\sect{2.2}) that in the one-shot game defection is the only evolutionary
stable strategy (ESS) \citep{Smith:82} \emph{if} each player interacts
with any other player. Such a population is called \emph{panmictic}, and
their dynamics can be predicted within a \emph{mean-field analysis}.
Interestingly, this picture changes if a \emph{spatial structure} of the
population is considered, i.e. if the interaction between agents is
\emph{locally restricted} to their neighbors \citep{Nowak:93,
  Nakamaru:97, Grim:97, Kirchkamp:00, Lindgren:94a, Oliphant:98,
  Vainstein:02}. Then, a stable \emph{spatial coexistence} between
cooperators and defectors becomes possible under certain conditions
\citep{Nowak:92b, Nowak:93, 
Oliphant:98}.

The current paper focuses on the spatial interaction of cooperators and
defectors both analytically and by means of computer simulations. We
consider an agent population placed on a square lattice and simulate
the dynamics of the population by means of a cellular automata defined in
\sect{2.1}. The interaction between the agents is modeled as a
\emph{n-person} game, i.e. $n$ agents interact simultaneously. In this
paper, a \emph{5-person game} is considered defined by the spatial
structure of each agent's four nearest neighbors. It is already known
that multi-person games produce a rather complex collective dynamics
\citep{Akimov:94, Dugatkin:89} which becomes even more difficult in the
spatial case. Therefore, in Sects. 3 and 4, we derive analytically
critical conditions for the invasion or the coexistence of cooperating
and defecting agents. In particular, we verify analytically the critical
parameters found by \citet{Nowak:93} by means of computer simulations.  A
very detailed investigation leads us to a phase diagram that allows to
distinguish between five different dynamic regimes for the spatial
evolution. Examples for these are demonstrated in Sect. 4 by means of
computer simulations. In Sect. 5 we present a modification of the model,
that to our knowledge for the first time investigates a true 5-person
game on the lattice. The results indicate that these modifications may
lead to an increase of cooperation in the spatial population.

\section{Model of Cooperation}
\label{2}
\subsection{Defining the cellular automaton}
\label{2.1}

We consider a model of two types of agents, cooperators $(C)$ and
defectors $(D)$. The disctinction is made by means of a variable
$\theta\in\{0,1\}$, where 1 refers to \emph{cooperation} and 0 refers to
\emph{defection}.  Whether an individual agent $i$ belongs to the
subpopulation of $C$ or $D$ is indicated by a variable, $\theta_{i}\in
\theta$. The total number of agents is constant, thus the global
frequency $f_{\theta}$ is defined as:
\begin{eqnarray}
  \label{nconst}
  N&=&\sum_{\theta}N_{\theta}=N_{0}+N_{1}= \mathrm{const.} \nonumber \\
 f_{\theta}&=&\frac{N_{\theta}}{N} \;; \quad f \equiv f_{1} = 1-f_{0}
\end{eqnarray}
In the following, the variable $f$ shall refer to the global frequency of
the cooperating agents.

For the spatial distribution of the agents let us consider a
two-dimensional cellular automaton (CA) (or a  two-dimensional square
lattice) consisting of
$N$ cells. 
Since the cells can be numbered consecutively, each cell is identified by
the index $i \in N$, that also refers to its spatial position. We note
that in most two-dimensional CA the position of the cells are identified
by $ij$ coordinates referring to the two dimensions, however in order to
define neighborhoods and to simplify the notations, we will use only
index $i$ for the spatial position. This implies that the description
presented below may also apply to one-dimensional CA provided
the neighborhoods are defined accordingly.

Each cell shall be occupied by one agent, thus each cell is
characterized by a discrete value $\theta_{i}\in\{0,1\}$ indicating whether
it is occupied by a cooperator or a defector. The spatiotemporal
distribution of the agents is then described by
\begin{equation}
  \label{vector}
  \mathbf{\Theta}=\{\theta_1,\theta_2,...,\theta_N\}
\end{equation}
Note that the state space $\Omega$ of all possible configurations is of
the order $2^{N}$.

\begin{figure}[htbp]
  \begin{center}
{\psset{unit=0.8}
    \begin{pspicture}(7,7)
\multirput(0.5,0.5)(1,0){7}{$\bullet$}   
\multirput(0.5,1.5)(1,0){3}{$\bullet$}
\multirput(4.5,1.5)(1,0){3}{$\bullet$}
\multirput(0.5,2.5)(1,0){2}{$\bullet$}
\multirput(5.5,2.5)(1,0){2}{$\bullet$}
\rput(0.5,3.5){$\bullet$}
\rput(6.5,3.5){$\bullet$}
\multirput(0.5,4.5)(1,0){2}{$\bullet$}
\multirput(5.5,4.5)(1,0){2}{$\bullet$}
\multirput(0.5,5.5)(1,0){3}{$\bullet$}
\multirput(4.5,5.5)(1,0){3}{$\bullet$}
\multirput(0.5,6.5)(1,0){7}{$\bullet$}  
\pspolygon[fillcolor=yellow,fillstyle=solid,linestyle=none]
(1,3)(2,3)(2,2)(3,2)(3,1)(4,1)(4,2)(5,2)(5,3)(6,3)(6,4)(5,4)(5,5)(4,5)
(4,6)(3,6)(3,5)(2,5)(2,4)(1,4)(1,3)
\pspolygon[fillcolor=green,fillstyle=solid,linestyle=none]
(2,3)(3,3)(3,2)(4,2)(4,3)(5,3)(5,4)(4,4)(4,5)(3,5)(3,4)(2,4)(2,3)
\pspolygon[fillcolor=white,fillstyle=solid,linestyle=none]
(3,3)(4,3)(4,4)(3,4)
      \rput(3.5,3.5){$\theta_{i}$}
      \rput(2.5,3.5){$\theta_{i_{1}}$}
       \rput(3.5,4.5){$\theta_{i_{2}}$}
     \rput(4.5,3.5){$\theta_{i_{3}}$}
      \rput(3.5,2.5){$\theta_{i_{4}}$}
      \rput(1.5,3.5){$\theta_{i_{5}}$}
       \rput(2.5,4.5){$\theta_{i_{6}}$}
     \rput(3.5,5.5){$\theta_{i_{7}}$}
      \rput(4.5,4.5){$\theta_{i_{8}}$}
      \rput(5.5,3.5){$\theta_{i_{9}}$}
       \rput(4.5,2.5){$\theta_{i_{10}}$}
     \rput(3.5,1.5){$\theta_{i_{11}}$}
  \rput(2.5,2.5){$\theta_{i_{12}}$}
    \end{pspicture}
}
  \end{center}
\caption[]{Sketch of the two-dimensional cellular automaton. It shows a
  neighborhood of size $n$ of cell $i$ where the neighbors are labeled by
  a second index $j=1,...,n-1$. The value $\theta_{i_{j}}\in \{0,1\}$
  indicates whether the cell is occupied either by a defector or a
  cooperator. The $m$ nearest neighbors are shown in darker gray, the $r$
  second nearest neighbors in lighter gray.
      \label{lattice}}
\end{figure}
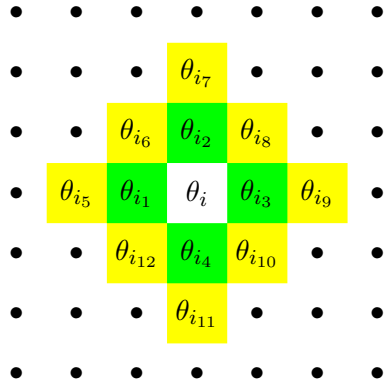
For the spatial evolution of $\bbox{\Theta}$ described in the following,
we have to consider the \emph{occupation distribution} of the \emph{local
  neighborhood} that surrounds each cell $i$. Let us define the size of a
neighborhood by $n$ (that also includes cell $i$), then the different
neighbors of $i$ are characterized by a second index $j=1,...,n-1$, where
the numbering starts with the nearest neighbors (cf.  \pic{lattice}).
Note that $\theta_{i}\equiv \theta_{i_{0}}$ (i.e $j=0$).
The number of \emph{nearest neighbors} of cell $i$ shall be denoted as
$m$, whereas the number of only the second nearest neighbors is denoted
as $r$. Obviously, $n=m+r+1$.  For further use, we define the local
occupation $\ul{\theta}_{i}$ of the nearest neighborhood (without cell
$i$) as:
\begin{equation}
  \label{occupat}
\ul{\theta}_{i}= \{\theta_{i_1},\theta_{i_2},...,\theta_{i_{m}}\}  
\end{equation}

\subsection{Rules of the Game}
\label{2.2}

In order to specify the interaction between the agents, we adopt a well
known prototype from game theory. It assumes that each agent $i$ has two
options to \emph{act} in a given situation, to \emph{cooperate} ($C$),
$\theta_{i}=1$, or to \emph{defect} ($D$), $\theta_{i}=0$. Playing with
agent $j$, the outcome of this interaction depends on the action chosen
by agent $i$, i.e. $C$ or $D$, \emph{without knowing} the action chosen
by the other agent participating in a particular game.  It is described
by a payoff matrix, which for the 2-person game has the following form:
\begin{center}
  \begin{equation}
    \label{2payoff}
    {\psset{unit=0.8}
      \begin{pspicture}(3,3)
        \rput(0,0){$\mathbf{D}$}
        \rput(1,0){0}
        \rput(2,0){T}
        \rput(3,0){P}
        \rput(0,1){$\mathbf{C}$}
        \rput(1,1){1}
        \rput(2,1){R}
        \rput(3,1){S}
        \rput(2,2){1}
        \rput(3,2){0}
        \rput(2,3){$\mathbf{C}$} 
        \rput(3,3){$\mathbf{D}$}
        \rput(1.2,2.2){$\theta_{j}$}
        \rput(0.8,1.8){$\theta_{i}$}
        \psline(0.5,2.5)(1.5,1.5)
        \psline(0.5,-0.3)(0.5,2.5)
        \psline(1.5,-0.3)(1.5,2.5)
        \psline(0.5,1.5)(3.3,1.5)
        \psline(0.5,2.5)(3.3,2.5)
        \psline(0.5,-0.3)(3.3,-0.3)
        \psline(3.3,-0.3)(3.3,2.5)
      \end{pspicture}}
  \end{equation}
\end{center}

Suppose, agent $i$ has chosen to cooperate, then its payoff is $R$ if the
other agent $j$ has also chosen to cooperate (without knowing about the
decision of agent $i$), but $S$ if agent $j$ defects. On the other hand,
if agent $i$ has chosen \emph{not} to cooperate, then it will receive the
payoff $T$ if agent $j$ cooperates, but $P$ if agent $j$ defects, too.

In a special class of games, the so-called \emph{Prisoner's Dilemma}
(PD), the payoffs have to fulfill the following two inequalities:
\begin{eqnarray}
\label{pd-ineq1}
T > R > P > S \\
\label{pd-ineq2}
2\,R > S+T
\end{eqnarray}
The known standard values are $T=5$, $R=3$, $P=1$, $S=0$. This means in a
cooperating environment, a defector will get the highest payoff.  From
this, the abbreviations for the different payoffs become clear: $T$ means
(T)emptation payoff for defecting in a cooperative environment, $S$ means
(S)ucker' payoff for cooperating in a defecting environment, $R$ means
(R)eward payoff for cooperating in a likewise environment, and $P$ means
(P)unishment payoff for defecting in a likewise environment.

In any \emph{one} round (or ``one-shot'') game, choosing action $D$
is unbeatable, because it rewards the higher payoff for agent $i$ whether
the opponent chooses $C$ or $D$. At the same time, the payoff for
\emph{both agents $i$ and $j$} is maximized when both cooperate. 

A simple analysis shows that defection is a so-called \emph{evolutionary
  stable strategy} (ESS) in a one-shot PD.  If any two agents are able to
interact and the number of cooperators and defectors in the population is
given by $N_1$ and $N_0=N-N_{1}$ respectively, then the expected average
payoff for cooperators will be $\bar{a}_{1} = (R \times N_1 + S \times
N_0)/N$.  Similarly the expected average payoff for defectors will be
$\bar{a}_{0}= (T\times N_1 + P \times N_0)/N$.  Since $T>R$ and $P>S$,
$\bar{a}_{0}$ is always larger than $\bar{a}_{1}$ for a given number
$N_{1}$, and pure defection would be optimal in a one-shot game. Even one
defector is sufficient to invade the complete population of $N-1$
cooperators.

This conclusion holds for a so called \emph{panmictic population} where
each agent interacts with all other $N-1$ agents. But in this paper, we
are mainly interested in the \emph{spatial effects} of the PD game.
Therefore, let us assume that each agent $i$ interacts only with the $m$
agents of its neighborhood, \eqn{occupat}.  In evolutionary game theory
this is called a $n$-person game, where $n=m+1=5$ in the given case.
Each game is played between $n$ players simultaneously, and the payoff
for each player depends on the number of cooperators in its group,
$s\leq n$.  Then the payoff matrix, \eqn{2payoff} has to be extended
according to the number of players, but it still has to fullfill the
following conditions according to the Prisoner's Dilemma rationale
\citep{Yao:94}:
\begin{enumerate}
\item Given a fixed number $s$ of cooperators in the group,
  defection pays always more than cooperation:
  \begin{equation}
    \label{cond1}
  a_{0}^{s} > a_{1}^{s} \;;\quad s=0,1,2,...,n-1   
  \end{equation}
  where $a_{0}^{s}$ is the payoff for each defector in the group of
  size $n$ and $a_{1}^{s}$ is the payoff for each cooperator.
\item The payoff increases for both cooperators and defectors, as number
  $s$ of cooperators in the group increases:
   \begin{equation}
     \label{cond2}
     a_{1}^{s} > a_{1}^{s-1} \;; \quad 
     a_{0}^{s} > a_{0}^{s-1} \;;\quad s=1,2,...,n-1
   \end{equation}
 \item The average payoff of the group increases with an increasing
   number of cooperators: 
   \begin{equation}
     \label{cond3}
     a_{1}^{s} s + a_{0}^{s} (n-s)>
     a_{1}^{s-1} (s-1) + a_{0}^{s-1} (n-s+1)
   \end{equation}
\end{enumerate}
There are various ways to fulfill the conditions
(\ref{cond1})-(\ref{cond3}). One simple method uses the concept of the
2-person game described above to calculate the payoff of an agent in the
$n$-person game from the accumulated payoffs of $(n-1)$ 2-person games.
I.e. for a neighborhood of 5, the 5-person game is decomposed into
2-person games, that may occur \emph{independently}, but
\emph{simultaneously} \citep{Lindgren:98}. This method will be
also used in the current paper. 

With respect to the spatial interaction in the neighborhood, we will
further assume that each agent interacts only with the $m$ nearest
neighbors \citep{Nowak:93}. In \sect{5}, we will also shortly discuss
the case where each agent in the neighborhood of size $n$ interacts with
every other agent in this neighborhood (which sometimes also involves
interaction with second nearest neighbors, as \pic{lattice} indicates).
These assuptions will allow us to further use the payoff matrix of the
2-person game, \eqn{2payoff}, and to reduce the interaction in the
neighborhood of size $n$ to $n-1$ simultaneous 2-person games played by
agent $i$.

In order to introduce a \emph{time scale}, we define a \emph{generation}
$G$ to be the time in which each agent has interacted with its $m$
nearest neighbors a given number of times, denotes as $n_{g}$. In our
case $n_{g}=1$ (which is called a \emph{one-shot} game), thus the total
number of interaction during each generation is roughly $N\,m/2$.  We
note here that $n_{g}>1$ can play an important role for the invasion of
cooperation into a heterogeneous population, \citep{Fogel:95, behera:02}.

We assume that during each generation the state $\theta_{i}$ of an agent
is not changed while it interacts with its neighbors simultaneously.
But after a generation is completed, $\theta_{i}$ can be changed based on
the following considerations: From the interaction with its neighbors,
each agent $i$ receives a \emph{payoff} $a_{i}$ that depends both on its
current state $\theta_{i}$, i.e. whether agent $i$ has cooperated or
not, and on the $\theta_{i_{j}}$ of its neighbors.  The fraction of
cooperators and defectors in the neighborhood of agent $i$ is given by:
\begin{equation}
  \label{sum}
z_{i}^{\theta} = \frac{1}{m}\sum_{j=1}^{m}
\delta_{\theta\theta_{i_{j}}} \;;\quad 
z_{i}^{(1-\theta)} = 1 - z_{i}^{\theta} \qquad (\theta\in \{0,1\} )
\end{equation}
where $\delta_{xy}$ means the Kronecker delta, which is 1 only for $x=y$
and zero otherwise.  According to the payoff matrix of \eqn{2payoff},
the payoff of agent $i$ is then defined as:
\begin{equation}
  \label{payoff-i}
  a_{i}(\theta_{i})= \delta_{1\theta_{i}}\, \Big[z_{i}^{1}\,R+z_{i}^{0}\,
  S\Big] 
+  \delta_{0\theta_{i}} \,\Big[z_{i}^{1}\,T+z_{i}^{0}\,P\Big]
\end{equation}
Note that the additive calculation of the payoff according to
\eqs{sum}{payoff-i} is based on the assumtion of the $m$ simultaneous, but
independent 2-person games played by agent $i$. 
 
The payoff $a_{i}$ is then compared to the payoffs $a_{i_{j}}$ of all
neighboring agents, in order to find the maximum payoff within the local
neighborhood during that generation, $\max{\{a_{i},a_{i_{j}}\}}$. If
agent $i$ has received the highest payoff, then it will keep its
$\theta_{i}$, i.e. it will \emph{continue} to either cooperate or to
defect. But if one of its neighbors $j$ has received the higher payoff,
then agent $i$ will adopt the behavior of the respective agent. This also
implies that it may keep its previous behavior if the more successful
agent has the same $\theta_{j}$. If 
\begin{equation}
  \label{star}
  {j}^{\star}= \arg\max\nolimits_{j=0,...,m}{a_{i_{j}}}
\end{equation}
defines the position of the agent that received the highest payoff in the
neighborhood, the update rule of the game can be concluded as follows:
\begin{equation}
  \label{update0}
 \theta_{i}(G+1)=\theta_{i_{j^{\star}}}(G)  
\end{equation}
which means specifically: 
\begin{equation}
  \label{update}
\theta_{i}(G+1)= \left \{
\begin{array}{cl}
\theta_{i}(G) & \quad \mbox{if} \;
\theta_{i}(G)= \theta_{i_{j^{\star}}}(G)\\ 
1-\theta_{i}(G) & \quad \mbox{otherwise} 
\end{array}
\right. 
\end{equation}
We note that the evolution of the system described by \eqn{update} is
completely \emph{deterministic}, results for stochastic CA have been
discussed in \citep{Durrett:99, fs-voter-02, muehlenb-hoens-02}. 

\subsection{First insights from the CA simulations}
\label{2.3}

Eventually, to give an impression of the spatial dynamics that arises
from the game desribed above, we have conducted a computer simulation of
the CA, using periodic boundary conditions and a simultaneous update. 
The global frequency of cooperators was initially set to $f(0)=0.5$.
The results are shown in the time series of \pic{fig:evolution} and the
time dependence of $f(G)$ in \pic{fig:freq}.
\begin{figure}[htbp]
\hspace{8pt}
\psfig{file=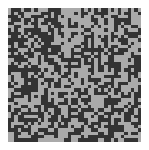,width=2.2cm}(0)\hspace{5pt}
\psfig{file=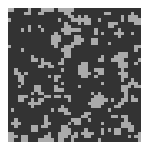,width=2.2cm}(1)\hspace{5pt}
\psfig{file=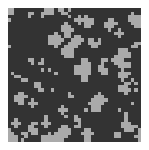,width=2.2cm}(2)\hspace{5pt}
\psfig{file=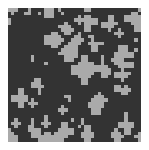,width=2.2cm}(3)

\hspace{8pt}
\psfig{file=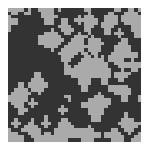,width=2.2cm}(4)\hspace{5pt}
\psfig{file=figures/gr5.ps,width=2.2cm}(5)\hspace{5pt}
\psfig{file=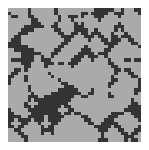,width=2.2cm}(10)\hspace{0.5pt}
\psfig{file=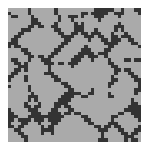,width=2.2cm}(20)
\caption{Time series of the spatial distribution of cooperators (grey)
  and defectors (black) on a CA of size $N=40\times 40$. The time is
  given by the numbers of generations in brackets.  Initial condition:
  $f(0)=0.5$, random spatial distribution of cooperators and defectors.
  Parameters for the payoff matrix, \eqn{2payoff}:
  $\{R;S;T;P\}=\{3.0;0.0;3.5;0.5\}$ (region A, see \sect{4.1}).
  \label{fig:evolution}}
\end{figure}
\begin{figure}[htbp]
\centerline{\psfig{file=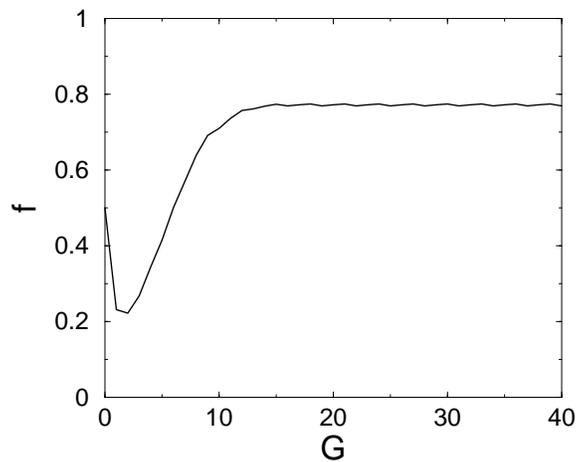,width=7.5cm}}
\caption{Global frequency of cooperators, $f$, \eqn{nconst}, vs. time
  (number of generations). 
  The data have been obtained from the time series of
  \pic{fig:evolution}.    \label{fig:freq}}
\end{figure}

From the computer simulations, we notice the following interesting
features of the spatial game: Starting with an initial random
distribution of cooperators and defectors, we observe the formation of
spatial domains dominated by either cooperators or defectors.  In the
very early stages of evolution, the cooperators concentrate in only a few
small clusters that are like islands in the sea of defectors.  But then
these cooperating clusters increase in size, i.e. the cooperators invade
into the domains of defectors, until in the long run they are the global
majority.  At all times, we observe a spatial \emph{coexistence} between
cooperators and defectors in different domains, instead of a complete
spatial separation into two domains. We further notice that in the long
run the spatial pattern becomes stationary (with very small periodic
changes that can be also noticed in \pic{fig:freq}).

In order to generalize the above results that have been obtained from
only a single run of the CA, we need to answer the following questions:
\begin{itemize}
\item Under what conditions will there be a spatial coexistence of
  cooperators and defectors, and when will we observe the extinction of
  either cooperators or defectors?
\item What does the variety and the size of the spatial domains depend
  on?
\item Will the dynamics always lead to stationary patterns, or will there
  be also non-stationary patterns in the long run?
\item How does the dynamics changes if a larger local neighborhood is
  considered in the update rule (i.e. if the second nearest neighbors are
  included, too)?
\end{itemize}
These questions will be investigated in detail in the following sections.
In particular, we will derive analytical results for the conditions that
need to be fulfilled by the \emph{payoff matrix}, \eqn{2payoff} in order
to find certain spatial patterns.

\section{Invasion vs. coexistence of  cooperating and defecting agents}
\label{3}
\subsection{Payoffs for local configurations}
\label{3.1}
As pointed out in the previous section, the change of agent $i$'s
``behavior'', $\theta_{i}\to(1-\theta_{i})$, depends on the payoff of the
agents in the immediate neighborhood, which in turn depends on the local
occupation, $\ul{\theta}_{i}$, \eqn{occupat}. In order to derive
analytical conditions for the transition of $\theta_{i}$, we thus have to
look more closely at the possible local occupation patters that shall in
the following be described by a term $K_{\theta}^{s}$. Here,
$\theta\in\{0,1\}$ describes whether the center cell is occupied by
either a defector or a cooperator, and $s \in \{0,1,2,3,4\}$ gives the total
number of \emph{cooperators} in the nearest neighborhood. Examples for
possible configurations in the neighborhood of a \emph{defector} are
shown in \eqn{config}:
\begin{equation}
  \label{config}
\begin{array}{ccc}
          &C& \\
          C&D&C\\
          &C& \\[4pt]
          &K_0^{4}&
        \end{array}\hfill
      \begin{array}{ccc}
          &C& \\
          C&D&D\\
          &C&\\[4pt]
          &K_0^{3}&
        \end{array}\hfill
      \begin{array}{ccc}
          &D& \\
          C&D&D\\
          &C&\\[4pt]
          &K_0^{2}&
        \end{array}\hfill
      \begin{array}{ccc}
          &D& \\
          D&D&D\\
          &C&\\[4pt]
          &K_0^{1}&
        \end{array}\hfill
       \begin{array}{ccc}
           &D& \\
           D&D&D\\
           &D&\\[4pt]
           &K_0^{0}&
         \end{array}
\end{equation}
Of course, for any given number $s\in \{1,2,3\}$, there are different
local occupation patterns possible that may result from exchanging the
positions of the agents in the neighborhood. But, as we have seen from
\eqn{payoff-i}, it is not really the \emph{position} of agents that
matters here, but only the number of cooperators and defectors in the
local neighborhood, \eqn{sum}. Therefore, $K_{0}^{s}$ provides sufficient
information about the local configuration. The possible configurations
for $K_{1}^{s}$, where a \emph{cooperator} occupies the center cell
are analogous to \eqn{config}.

For each possible realization of $K_{\theta}^{s}$, the respective payoff
of the agent in the center, $a_{\theta}^{s}$ can be calculated according
to \eqn{payoff-i}. The results are given by \eqn{payoff-both} for
cooperators and defectors both in terms of an analytical expression and
for a particular realization of the payoff matrix, \eqn{2payoff}, 
$\{R;S;T;P\}=\{3;0;5;1\}$ (which are the ``classical'' values):
\begin{equation}
  \label{payoff-both}
  \begin{array}{lcl}
\multicolumn{3}{c}{\mathrm{cooperator}} \\[3pt]
a_1^{4}=& R&= 3.0\\[3pt]
a_1^{3}=& \frac{\D 3R+S}{\D 4}&=2.25\\[3pt]
a_1^{2}=& \frac{\D 2R+2S}{\D 4}&=1.5\\[3pt]
a_1^{1}=& \frac{\D R+3S}{\D 4}&=0.75\\[3pt]
a_1^{0}=& S&=0.0
\end{array} \hfill
  \begin{array}{lcl}
\multicolumn{3}{c}{\mathrm{defector}} \\[3pt]
a_0^{4}=&T&=5.0\\
a_0^{3}=&\frac{\D 3T+P}{\D 4}&=4.0 \\[3pt]
a_0^{2}=&\frac{\D 2T+2P}{\D 4}&=3.0\\[3pt]
a_0^{1}=&\frac{\D T+3P}{\D 4}&=2.0\\[3pt]
a_{0}^{0}=&P&=1.0
\end{array}
\end{equation}
From \eqn{payoff-both}, we notice that the payoff $a_{\theta}^{s}$
increases with the number of cooperators $s$ in the neighborhood, i.e.
for a particular $\theta$
\begin{equation}
  \label{neq}
  a_{\theta}^{s}>a_{\theta}^{s-1} \quad(s \in \{1,...,4\})
\end{equation}
yields. Further, taking the relations of \eqs{pd-ineq1}{pd-ineq2} into
account, it is obvious that always 
\begin{eqnarray}
  \label{neq2}
   a_{0}^{4}&=&\max{\{a_{0}^{s}, a_{1}^{s}\}} \nonumber \\
   a_{1}^{0}&=&\min{\{a_{0}^{s}, a_{1}^{s}\}}\\  
   a_{0}^{s}&>&a_{1}^{s} \quad 
(s\in\{0,1,...,4\}) \nonumber
\end{eqnarray}
Comparing \eqs{neq}{neq2} with the conditions
(\ref{cond1})-(\ref{cond3}), we can also verify that the payoffs in the
current form fulfill the general requiremements for the n-person PD.
Based on these general considerations, in the following we discuss the
different conditions that arise \emph{in the spatial case} for the
invasion or the coexistence of cooperators and defectors.

\subsection{Conditions for invasion and coexistence}
\label{3.2}
Basically, the update rule of \eqn{update} determines whether a cell is
``invaded'' by a particular subpopulation (i.e. whether the agent adopts
the respective behavior). It depends on the local payoff received in
comparison with the payoffs of the neighboring sites.  In order to find
the conditions for invasion and coexistence, we need to concentrate on
the \emph{border region} between the domains of cooperators and defectors
that is sketched in \pic{border}. $K_{1}^{4}$ describes the local
configuration \emph{inside} the domain of cooperators, while $K_{0}^{0}$
describes the local configuration \emph{inside} the domain of defectors.
Further, the possible configurations that can be found only in the border
region are given. The local configurations $K_{1}^{1}$ and $K_{1}^{0}$
are not listed there because they are completely unstable.  These
configurations receive the lowest payoff and therefore, if they initially
exist, vanish within the next generation and do not need to be taken into
account for the discussion of the further evolution.
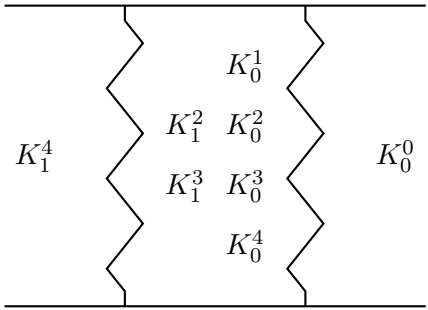
\begin{figure}[htbp]
  \begin{center}
{\psset{unit=0.8}
    \begin{pspicture}(7,5)
\psline(0.0,0.0)(7.0,0.0)
\psline(0.0,5.0)(7.0,5.0)
\pszigzag[coilarm=0.01,coilwidth=0.6,coilheight=2.5](2.0,0.0)(2.0,5.0)
\rput(0.5,2.5){$K_{1}^{4}$}
\rput(3.0,2.0){$K_{1}^{3}$}
\rput(3.0,3.0){$K_{1}^{2}$}
\rput(4.0,1.0){$K_{0}^{4}$}
\rput(4.0,2.0){$K_{0}^{3}$}
\rput(4.0,3.0){$K_{0}^{2}$}
\rput(4.0,4.0){$K_{0}^{1}$}
\rput(6.5,2.5){$K_{0}^{0}$}
\pszigzag[coilarm=0.01,coilwidth=0.6,coilheight=2.5](5.0,0.0)(5.0,5.0)
    \end{pspicture}}
\end{center}
\caption{Sketch of the possible neighboring configurations in a border
  region between domains of cooperators ($K_{1}^{4}$) and defectors
  ($K_{0}^{0}$).
  \label{border} }
\end{figure}

In order to elucidate the dynamics at the border, we will separately
discuss the two possible cases: (i) invasion of configurations
$K_{0}^{s}$ ``owned'' by \emph{defectors} into neighboring configurations
$K_{1}^{s}$ ``owned'' by \emph{cooperators}, and (ii) the reverse case.
Note that the cooperator $(1)$ and the defector $(0)$ are always on
adjacent sites. %
The invasion process, i.e. the occupation of the center cell by an agent
of the opposite subpopulation shall be indicated by an arrow $\RA$.  The
results for case (i) are listed in Table \ref{tab-defect} and are
explained below.
\begin{table}[htbp]
    \caption{Invasion of configurations
      $K_{0}^{s}$ ``owned'' by defectors into configurations $K_{1}^{s}$
      ``owned'' by cooperators in the border region, \pic{border}.}
    \begin{center}
   {\begin{tabular}[c]{ccl}
\hline \hline
configuration & necessary conditions & stationary state \\ \hline
$K_{0}^{1}\RA K_{1}^{2}$ & $a_{0}^{1}>a_{1}^{4}$ & complete invasion \\
$K_{0}^{1}\RA K_{1}^{3}$ & & \\ \hline
$K_{0}^{2}\RA K_{1}^{2}$ & $a_{0}^{2}>a_{1}^{4}$ & unstable  \\ 
$K_{0}^{2}\RA K_{1}^{3}$ & $a_{1}^{4}>a_{0}^{1}$ & coexistence (d) \\ \hline
$K_{0}^{3}\RA K_{1}^{2}$ & $a_{0}^{3}>a_{1}^{4}$ &   \\
$K_{0}^{3}\RA K_{1}^{3}$ & $a_{0}^{2}<a_{1}^{3}$ & coexistence (b) \\ 
& $a_{0}^{1}<a_{1}^{2}$ &   \\
\hline \hline
    \end{tabular}} 
    \end{center}
    \label{tab-defect}
\end{table}

The defective behavior can invade the whole spatial population without
any exception (in a deterministic case), if the \emph{lowest} possible
payoff for defectors, $a_{0}^{1}$, is \emph{larger} than the
\emph{highest} possible payoff for collaborators, $a_{1}^{4}$. Actually,
the payoff $a_{0}^{0}$ does not matter here, because in the respective
configuration $K_{0}^{0}$, there is no cooperator left to adopt the
defective strategy.  Because always
$a_{0}^{4}>a_{0}^{3}>a_{0}^{2}>a_{0}^{1}$, \eqn{neq}, this leads to the
necessary condition for the complete invasion of defectors into the
domain of cooperators:
\begin{equation}
  \label{invas}
  a_{0}^{1}>a_{1}^{4}
\end{equation}
The two remaining cases that lead to \emph{coexistence} will be explained
together with the next table, as well as the special case of
$K_{0}^{4}$. 

For the case (ii), invasion of configurations $K_{1}^{s}$ ``owned'' by
\emph{cooperators} into configurations $K_{0}^{s}$ ``owned'' by
\emph{defectors}, the results for the possible cases are listed in Table
\ref{tab-coop}.
\begin{table}[htbp]
    \caption{Invasion of configurations
      $K_{1}^{s}$ ``owned'' by cooperators into configurations $K_{0}^{s}$
      ``owned'' by defectors in the border region, \pic{border}.}
    \begin{center}
   {
         \begin{tabular}[c]{ccl}
\hline \hline
configuration & necessary conditions & stationary state \\ \hline
$K_{1}^{2}\RA K_{0}^{1}$ & $a_{1}^{2}>a_{0}^{1}$ & coexistence (a) \\
$K_{1}^{3}\RA K_{0}^{1}$ & $a_{1}^{3}>a_{0}^{2}$ & C majority \\ 
$K_{1}^{3}\RA K_{0}^{2}$ & $a_{1}^{4}>a_{0}^{3}$ & stationary domains 
\\ \hline
$K_{1}^{2}\RA K_{0}^{1}$ & $a_{1}^{2}>a_{0}^{1}$ & coexistence (b) \\
$K_{1}^{3}\RA K_{0}^{1}$ & $a_{1}^{3}>a_{0}^{2}$ & C minority \\ 
$K_{1}^{3}\RA K_{0}^{2}$ & $a_{1}^{4}<a_{0}^{3}$ & spatial chaos 
\\ \hline
 & $a_{0}^{2} >  a_{1}^{3} > a_{0}^{1}$ & coexistence (c) \\
$K_{1}^{3}\RA K_{0}^{1}$ & $a_{0}^{3} >  a_{1}^{4} > a_{0}^{2}$  
& C minority      \\   
 & & small clusters 
\\ \hline
$K_{1}^{2}\RA K_{0}^{2}$ & &  \\ 
$K_{1}^{2}\RA K_{0}^{3}$ & not possible & \\ 
$K_{1}^{2}\RA K_{0}^{4}$ &  & \\
$K_{1}^{3}\RA K_{0}^{3}$ &  & \\
$K_{1}^{3}\RA K_{0}^{4}$ & & \\ \hline \hline
    \end{tabular} 
} 
    \end{center}
   \label{tab-coop}
\end{table}
First, we notice that a \emph{complete invasion} of cooperators into the
whole population is not possible, because the necessary condition
$a_{1}^{1}>a_{0}^{4}$, i.e. the \emph{lowest} possible payoff for
cooperators is \emph{larger} than the \emph{highest} possible payoff for
defectors, is never fulfilled. Further, from \eqn{neq2} we realize that
configuration $K_{0}^{s}$ cannot be invaded by $K_{1}^{s}$ because the
former payoff is always higher. But configuration $K_{1}^{s}$ could
always invade a neighboring configuration $K_{0}^{s-1}$ \emph{if} the
condition
\begin{equation}
  \label{invas2}
  a_{1}^{s}>a_{0}^{s-1} \quad (s \in\{2,3,4\})
\end{equation}
would be satisfied. As \eqn{payoff-both} shows, this is not always the
case, but provided suitable values of $T$ and $P$, it can be possible.
Thus, $K_{1}^{2}$ could invade a neighboring configuration $K_{0}^{1}$, 
and $K_{1}^{3}$ could invade both $K_{0}^{2}$ and $K_{0}^{1}$. 
However, we recall that $K_{1}^{3}$ and $K_{1}^{2}$ cannot invade
$K_{0}^{3}$ 
because of \eqn{neq2}. This means that the invasion of cooperation
necessarily stops at some point.

On the other hand, $K_{0}^{3}$ and could invade $K_{1}^{3}$ and
$K_{1}^{2}$ because of \eqn{neq2}.  But this kind of invasion does not
happen if \eqn{invas2} is satisfied.  This can be explained by looking at
\pic{explain} where the payoffs in the border region are shown under the
assumption that \eqn{invas2} is valid. In the example shown, the payoff
of configuration $K_{0}^{3}$ is higher than $K_{1}^{2}$ or $K_{1}^{3}$
because of \eqn{neq2}. But because of \eqn{invas2},
$a_{1}^{4}>a_{0}^{3}$, thus the invasion of $K_{0}^{3}$ cannot take
place, i.e. the border remains at its current position.  A similar
explanation holds for the remaining case $K_{0}^{2}$.
\begin{figure}[htbp]
  \begin{center}
{\psset{unit=0.6}
    \begin{pspicture}(9,5)
\psline(0.5,0.0)(0.5,5.0)
\psline(0.5,0.0)(9.5,0.0)
\psline(1.0,4.5)(3.0,4.5)(3.0,1.5)(5.0,1.5)(5.0,3.5)(7.0,3.5)(7.0,0.5)(9.0,0.5)  
\rput(2.0,5.0){$a_{1}^{4}$}
\rput(4.0,1.0){$a_{1}^{2}$ or $a_{1}^{3}$}
\rput(6.0,4.0){$a_{0}^{3}$}
\rput(8.0,1.0){$a_{0}^{0}$}
\rput{90}(0.0,2.5){\large payoff}
    \end{pspicture}}
\end{center}
\caption{Payoffs in the border region between domains of cooperators
  ($K_{1}^{4}$) and defectors ($K_{0}^{0}$) with the assumption that
  \eqn{invas2} is valid. It explains that configuration $K_{0}^{3}$
  cannot invade neighboring configurations $K_{1}^{2}$ or $K_{1}^{3}$
  because of the higher payoff of configuration $K_{1}^{4}$ that
  ``backs'' them from the other side.
 \label{explain} }
\end{figure}
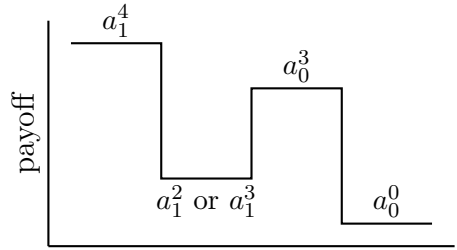

In conclusion, one (strong) condition for \emph{coexistence} is given by
\eqn{invas2}. For $s\in \{2,3,4\}$ this means basically three different
inequalities. If all of them are satisfied, the cooperators become the
global \emph{majority}, because they are always allowed to invade
$K_{0}^{1}$, $K_{0}^{2}$, whereas defectors in $K_{0}^{3}$ who could
possibly invade the cooperation domain are stopped, as explained above.
We could relax the condition of \eqn{invas2} that causes this stop. So,
let us assume the case that only \emph{two} of the coexistence
inequalities are satisfied, i.e.
\begin{equation}
  \label{invas3}
  a_{1}^{s}>a_{0}^{s-1} \;\; (s \in\{2,3\}) \;; \quad
a_{1}^{4}<a_{0}^{3}
\end{equation}
Then, the defector configuration $K_{0}^{3}$ can invade the cooperation
configurations $K_{1}^{2}$, $K_{1}^{3}$, which eventually leads to the
domination of defectors, i.e. the cooperators become a global
\emph{minority}. We note that in this case we will not find stable
coexisting domains, but a nonstationary pattern that appears to be
\emph{spatial chaos}, as also shown in the next section. 

If the strong condition for coexistence, \eqn{invas2} is not satisfied, a
weaker condition for coexistence could still hold true:
\begin{equation}
  \label{invas4}
  a_{0}^{s-1}>  a_{1}^{s} > a_{0}^{s-2}  \;\; (s \in\{3,4\})
\end{equation}
This condition would lead to an even greater domination of defectors,
because only the configuration $K_{0}^{1}$ can be invaded by
cooperators, while all other ones are protected from this. 
In this case, as we will also show by means of computer simulations, the
cooperators will only survive in very small clusters. 

A very special case arises if none of the above coexistence inequalities
are satisfied, but instead either one of the following conditions holds: 
\begin{eqnarray}
  \label{invas5}
a_{0}^{1}< a_{1}^{4}<a_{0}^{2}   \\
 \label{invas5b}
a_{0}^{2}<a_{1}^{4} \;;\quad a_{1}^{3}< a_{0}^{1} \\
 \label{invas5c}
a_{0}^{3}<a_{1}^{4} \;;\quad a_{1}^{2}< a_{0}^{1} < a_{1}^{3} 
\end{eqnarray}
In this case, \emph{if} we start from the special initial condition of
only two domains separated by a planar interface, we find that this is
also a stable configuration. But any deviation from this initial
condition will eventually lead to an invasion of defectors. In
particular, if we start from a random initial distribution of cooperators
and defectors, \emph{invasion} of defectors will always take place under
one of the conditions definded by eqs. (\ref{invas5})-(\ref{invas5c}).
Therefore, this case has been named \emph{unstable coexistence} (d) in
Table \ref{tab-defect}.  It denotes the transition from the case of
complete invasion of defectors to the case of stable coexistence between
cooperators and defectors.

The only remaining case to be explained is $K_{0}^{4}$ in the border
region, which arises if cooperation invades into the domain of defection
from different sides.  Then, it can happen that a single defecting agent
is trapped within the domain of cooperators. Because of
$a_{0}^{4}>a_{1}^{4}$, defection will invade all neighboring sites during
the next generation, this way leading to a configuration $K_{0}^{0}$ in
the border region. This could lead to a growth of the domain of defectors
if the conditions for complete invasion, \eqn{invas} is satisfied (see
also Table \ref{tab-defect}). But if these conditions are not satisfied,
there are various possibilities dependent on whether three, two or one of
the inequalities for coexistence are satisfied (see also Table
\ref{tab-coop}).  $K_{0}^{0}$ is surrounded by other configurations
$K_{0}^{1}$, $K_{0}^{2}$ or $K_{0}^{3}$ in the larger neighborhood.  If
\eqn{invas2} is valid and only $K_{0}^{1}$, $K_{0}^{2}$ are present, then
in the next generation they will be occupied by cooperators, as explained
above, leaving the defector in the central cell in its trapped situation,
again.  So, a cycle of $K_{0}^{4}$ and $K_{0}^{0}$ alternating is
created. If $K_{0}^{3}$ is present, this will create a stable border
between the spatial domains of cooperators, this way leading to
coexistence as explained above, too.  If \eqn{invas3} is valid and
$K_{0}^{3}$ is present, this will lead to the split-up of the domains of
cooperators.  If \eqn{invas4} is valid and $K_{0}^{3}$ is present, this
will lead to a growth of the defector domain.

In the following section, we will apply the analytical conditions derived
for the invasion and coexistence of cooperators and defectors to find out
critical values of the payoff matrix that may lead to the predicted
spatial patterns.

\section{Critical payoff values for invasion and coexistence}
\label{4}
\subsection{Derivation of a phase diagram}
\label{4.1}
The investigations of the previous section have provided us with a set of
inequalities for the payoffs $a_{\theta}^{s}$, in order to find certain
patterns of invasion and coexistence of cooperators and defectors.  The
possible payoffs $a_{\theta}^{s}$, \eqn{payoff-both} depend on the set of
variables $\{R;S;T;P\}$ of the payoff matrix, \eqn{2payoff}. In the
following, we fix the values for $R=3$ and $S=0$, i.e. the payoffs for a
\emph{cooperating} agent $i$, to the standard values, and this way derive
the critical conditions for the payoffs for a \emph{defecting} agent, $T$
and $P$. With the known payoffs $a_{\theta}^{s}$, \eqn{payoff-both}, we
find from the different inequalities (\ref{invas2}) the following set of
conditions:
\begin{eqnarray}
  \label{a-f}
 \mathrm{(a)} \quad a_{1}^{4} \lessgtr a_{0}^{3}  \quad \RA \quad & 
3T+P &=12 \nonumber \\
\mathrm{(b)} \quad a_{1}^{3} \lessgtr a_{0}^{2}  \quad \RA \quad & 
2T+2P &=9 \nonumber \\
\mathrm{(c)} \quad a_{1}^{2} \lessgtr a_{0}^{1}  \quad \RA \quad & 
T+3P& =6 \\
\mathrm{(d)} \quad a_{1}^{4} \lessgtr a_{0}^{2}  \quad \RA \quad & 
2T+2P&=12 \nonumber \\
\mathrm{(e)} \quad a_{1}^{3} \lessgtr a_{0}^{1}  \quad \RA \quad & 
T+3P&=9 \nonumber \\
\mathrm{(f)} \quad a_{1}^{4} \lessgtr a_{0}^{1}  \quad \RA \quad & 
T+3P& =12 \nonumber
\end{eqnarray}
Additionally, we always have 
\begin{equation}
  \label{tp}
3\leq T\leq 6 \;; \quad 0\leq P \leq 3  
\end{equation}
because of the in\eqs{pd-ineq1}{pd-ineq2}. With these restrictions, we
are able to plot a phase diagram in the $\{P,T\}$ parameter space,
\pic{phase} where the conditions (\ref{a-f}) mark the different
boundaries between the phases. The labels \textbf{A}-\textbf{E} in
\pic{phase} denote the parameter regions associated with the different
stationary patterns and shall be explained in detail in the following.
\begin{figure}[htbp]
\begin{center}
{\psset{unit=2.0}
    \begin{pspicture}(0,1.5)(4,6.5)
\pspolygon[fillcolor=red,fillstyle=solid,linestyle=none]
(0,3)(1,3)(0.75,3.75)(0,4)  
\pspolygon[fillcolor=cyan,fillstyle=solid,linestyle=none]
(0,4)(0.75,3.75)(0,4.5)  
\pspolygon[fillcolor=green,fillstyle=solid,linestyle=none]
(0,4.5)(0.75,3.75)(1.875,3.375)(1.5,4.5)(0,6)  
\pspolygon[fillcolor=yellow,fillstyle=solid,linestyle=none]
(0.75,3.75)(1,3)(3,3)(2,6)(0,6)(1.5,4.5)(1.875,3.375)  
\pspolygon[fillcolor=magenta,fillstyle=solid,linestyle=none]
(3,3)(2,6)(3,6)  
\psgrid[gridlabels=0](0,3)(3,6)
\psline[linewidth=3pt,linestyle=solid](0,3)(3.5,3)
\psline[linewidth=3pt,linestyle=solid](3,2.5)(3,6.5)
\psline[linewidth=3pt,linestyle=solid](0,6)(3.5,6)
\psline[linewidth=2pt,linestyle=solid](0,4)(3.5,2.8333) 
\rput(1.2,3.7){\large a}
\psline[linewidth=2pt,linestyle=solid](0,4.5)(2,2.5)    
\rput(1.2,3.2){\large b}
\psline[linewidth=2pt,linestyle=solid](0,6)(1.1666,2.5) 
\rput(0.8,3.2){\large c}
\psline[linewidth=2pt,linestyle=solid](0,6)(3.5,2.5)    
\rput(1.2,4.6){\large d}
\psline[linewidth=2pt,linestyle=solid](0.8333,6.5)(2.1666,2.5) 
\rput(1.2,5.7){\large e}
\psline[linewidth=2pt,linestyle=solid](1.8333,6.5)(3.1666,2.5) 
\rput(2.2,5.7){\large f}
\rput(0.4,3.4){\large \textbf{A}}
\rput(0.2,4.1){\large \textbf{B}}
\rput(0.8,4.5){\large \textbf{C}}
\rput(1.6,5.3){\large \textbf{D}}
\rput(2.6,5.3){\large \textbf{E}}
\psaxes[Ox=0,Oy=2,ticks=all,labels=all,showorigin=false,axesstyle=axes](0,2)(3.5,6.5)
\rput(1.5,1.5){\Large $\mathbf{P}$}
\rput{90}(-0.5,4.5){\Large $\mathbf{T}$}
\end{pspicture}
}
\end{center}
\caption{Phase diagram in the $\{T,P\}$ parameter space for fixed values
  of $R=3$ and $S=0$ of the payoff matrix, \eqn{2payoff}. The different
  lines result from \eqn{a-f} ($a-f$) and \eqn{tp}. The areas $A-E$ (in
  different gray scales) indicate parameter regions that lead to
  particular spatial patterns in the stationary limit. For a detailed
  explanation see text. Note that the ``classical'' parameter values
  $\{R;S;T;P\}=\{3;0;5;1\}$ are just on the border between regions $C$
  and $D$. \label{phase}}
\end{figure}
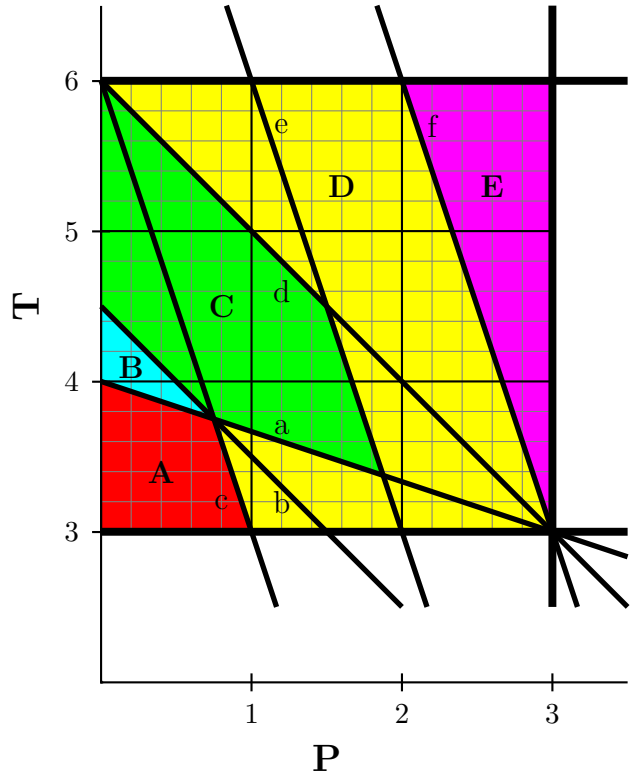

\subsection{Complete invasion of defectors (region E)}
\label{4.2}
For the complete invasion of defectors into the domain of cooperators,
the necessary condition of  
\eqn{invas} applies, i.e.
\begin{equation}  
\label{p-invas}
  \frac{T+3P}{4} > R
\end{equation}
With the fixed values for $R$ and $P$ and \eqn{tp} we find
\begin{equation}
  \label{p-invas2}
 12 - 3 P < T < 6 \quad \mathrm{if} \quad  2 < P <3 
\end{equation}
This condition defines region \textbf{E} in the parameter space
shown in \pic{phase}. Thus, for the appropriate $\{T,P\}$ values the
stationary state is always entirely dominated by the defectors
regardless of the initial conditions, therefore no further computer
simulations are presented here. 

\subsection{Coexistence with a majority of cooperators (region A)}
\label{4.3}
Coexistence between cooperators and defectors becomes possible if
\eqn{invas2} is fulfilled. As we have shown in \sect{3}, this leads to
different cases dependent on how many inequalities of \eqn{invas2} are
satisfied. If all of them are fulfilled, this leads to the conditions: 
\begin{equation}
\label{eq:constr3}
\frac{2R+2S}{4} > \frac{T+3P}{4} \;; \quad
\frac{3R+S}{4} > \frac{2T+2P}{4} \;; \quad
R> \frac{3T+P}{4}
\end{equation}
With the fixed values for $R$ and $S$ we find:
\begin{equation}
\label{eq:s31}
T+3P<6 \;; \quad 
2T+2P<9 \;; \quad
3T+P<12 
\end{equation}
where additionally \eqn{tp} applies. The solution is then given by:
\begin{equation}
\label{eq:s32}
\begin{array}{rclcrcl}
3 &<T<& 4-\frac{\D P}{\D 3} & \quad \mathrm{if} \quad & 0 &<P<& 0.75  \\
3 & <T<& 6- 3P & \quad \mathrm{if} \quad & 0.75 &<P<& 1.0 
\end{array}
\end{equation}
These two conditions define region \textbf{A} of \pic{phase}.  \Eqn{a-A}
gives the respective payoffs $a_{\theta}^{s}$, \eqn{payoff-both} for the
choosen set of parameters, $\{R;S;T;P\}=\{3.0;0.0;3.5;0.5\}$. The
$a_{\theta}^{s}$ can be used to verify the general considerations of
\sect{3}.  They are printed here to simplify the
comparison.
\begin{equation}
  \label{a-A}
  \begin{array}{ccccccccccc}
\quad a_{1}^{4} \quad & \quad a_{1}^{3} \quad & \quad a_{1}^{2} \quad &
\quad a_{1}^{1} \quad & \quad a_{1}^{0} \quad & \; \; & 
\quad a_{0}^{4} \quad & \quad a_{0}^{3} \quad & \quad a_{0}^{2} \quad & 
\quad a_{0}^{1} \quad & \quad a_{0}^{0} \quad \\
3.0 & 2.25 & 1.5 & 0.75 & 0.0 & & 3.5 & 2.75 & 2.0 & 1.25 & 0.5 
  \end{array}
\end{equation}
Note that the $a_{1}^{s}$ are the same as in the ``classical'' case,
\eqn{payoff-both} since they only depend on the values of $R$ and $S$
that are not changed.  For the appropriate $\{T,P\}$ values the
stationary state always shows a coexistence between cooperators and
defectors where the cooperators are a \emph{majority}. Note, that this
parameter region is just opposite to the region \textbf{E}, \pic{phase}
where the stationary state is characterized by an \emph{extinction} of
the cooperators.

The coexistence between cooperators and defectors shall be also
demonstrated by means of computer simulations. To elucidate the dynamics,
we have choosen two \emph{different initial conditions} for the
simulations, shown in \pic{ini}. The left part of \pic{ini} shows a
regular situation of only one cluster of cooperators is a sea of
defectors, i.e. the cooperators are the absolute \emph{minority}.  For
this initial condition, \pic{cluster} shows the evolution of the global
frequency for the different parameter regions \textbf{A}, \textbf{B} and
\textbf{C}.  The right part of \pic{ini} shows a random initial
distribution of cooperators and defectors with the same initial
frequency, $f(0)=0.5$. In order to study the influence of the initial
frequency on the long-term dynamics, we will also choose random
distributions with different values of $f(0)$.
\begin{figure}[htbp]
\centerline{
  \psfig{file=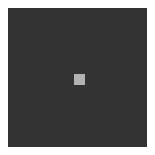,width=2.2cm}\hspace{25pt}
  \psfig{file=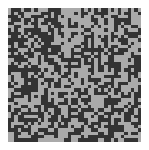,width=2.2cm}
}
\caption{Initial spatial distributions of cooperators (grey) and
  defectors (black) on a CA of size $N=40\times 40$, used for the
  computer simulations: (left) One cluster of 9 cooperators,
  $f(0)=0.0056$, (right) random spatial distribution of cooperators and
  defectors, shown for $f(0)=0.5$.
    \label{ini}}
\end{figure}
\begin{figure}[htbp]
 \hspace{8pt}
\psfig{file=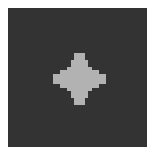,width=2.2cm}(5)\hspace{5pt}
\psfig{file=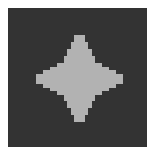,width=2.2cm}(10)\hspace{0.5pt}
\psfig{file=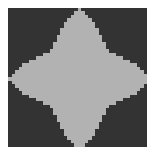,width=2.2cm}(20)\hspace{0.5pt}
\psfig{file=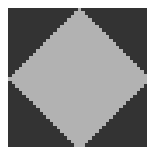,width=2.2cm}(30)\hspace{5pt}
    \caption{Time series of the spatial distribution of cooperators (grey)
      and defectors (black) on a CA of size $N=40\times 40$. The time is
      given by the numbers of generations in brackets.  Initial
      condition: One cluster of 9 cooperators, \pic{ini}(left).
      Parameters for the payoff matrix, \eqn{2payoff}:
      $\{R;S;T;P\}=\{3.0;0.0;3.5;0.5\}$ (region A).
    \label{square}}
\end{figure}
\begin{figure}[htbp]
\centerline{\psfig{file=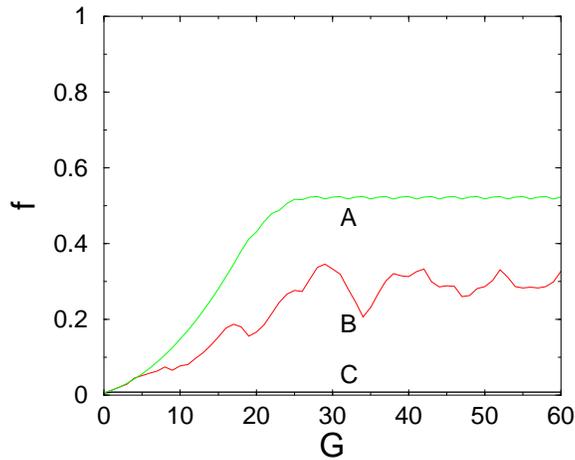,width=7.5cm}}
 \caption{Global frequency of cooperators, $f$, \eqn{nconst}, vs. time
   (number of generations).  Initial condition: One cluster of 9
   cooperators, \pic{ini}(left), $f(0)=0.0056$.  Parameters for the
   payoff matrix, \eqn{2payoff}, $\{R;S;T;P\}$: (A) $\{3.0;0.0;3.5;0.5\}$
   (region A), (B) $\{3.0;0.0;3.9;0.5\}$ (region B), (C)
   $\{3.0;0.0;4.5;1.0\}$ (region C). 
   \label{cluster} }
\end{figure}
\begin{figure}[htbp]
\centerline{
  \psfig{file=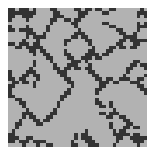,width=2.2cm}\hspace{15pt}
  \psfig{file=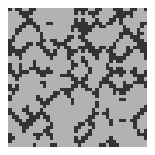,width=2.2cm}\hspace{15pt}
  \psfig{file=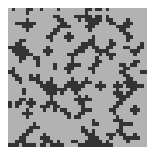,width=2.2cm}\hspace{15pt}
}
\caption{Final (steady state) spatial distribution of cooperators (grey) and
  defectors (black) on a CA of size $N=40\times 40$. Initial condition:
  random spatial distribution of cooperators and defectors, left:
  $f(0)=0.5$, middle: $f(0)=0.75$, right: $f(0)=0.9$. Parameters:
  $\{R;S;T;P\}=\{3.0;0.0;3.5;0.5\}$ (region A).
    \label{fig:case5}}
\end{figure}
\begin{figure}[htbp]
  \centerline{\psfig{file=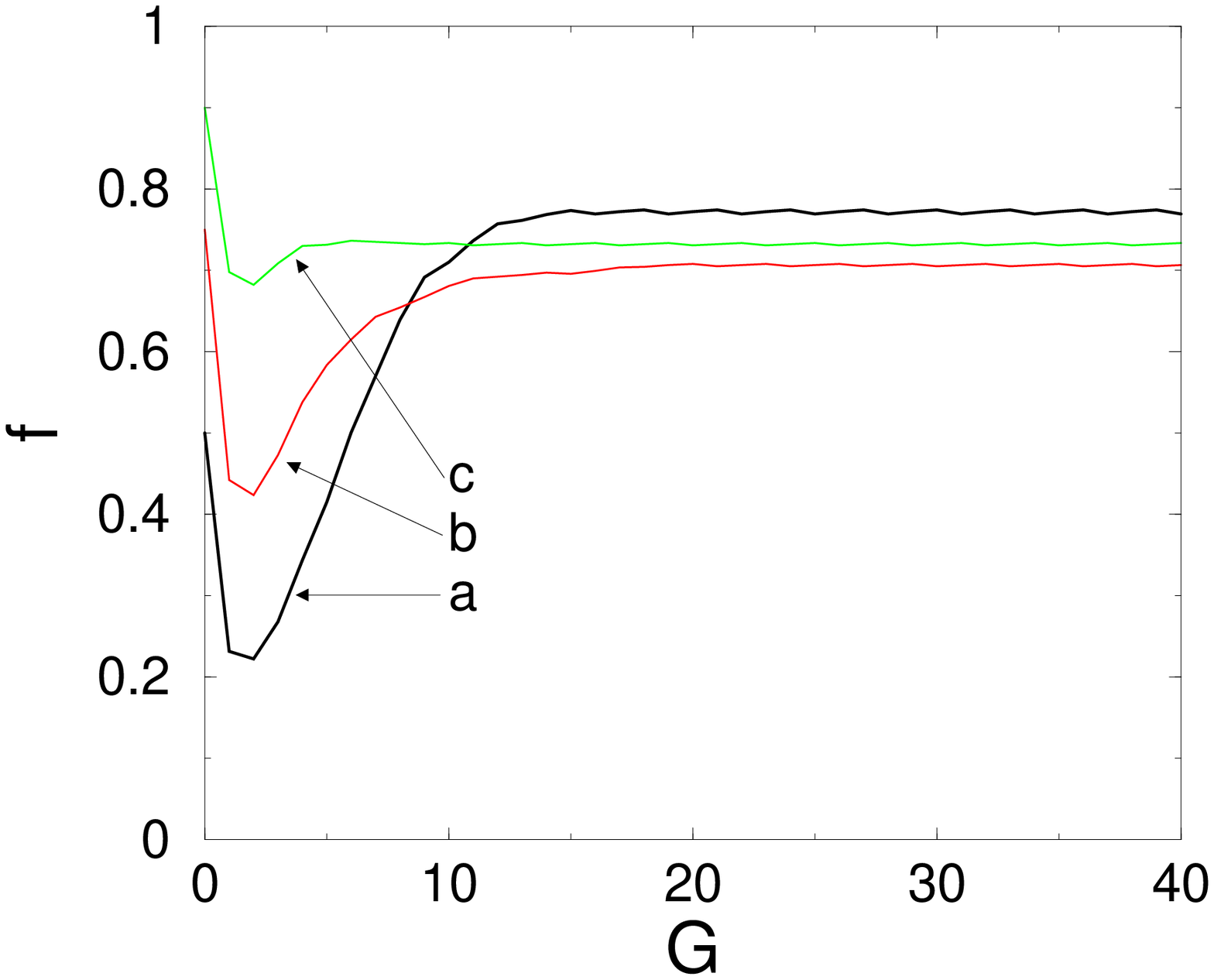,width=7.5cm}}
    \caption{Global frequency of cooperators, $f$,
      \eqn{nconst}, vs. time (number of generations).  The data have been
      obtained from simulations with three different initial frequencies:
      (a) $f(0)=0.5$, (b) $f(0)=0.75$, (c) $f(0)=0.9$. See also
      \pic{fig:case5} for the final spatial distributions. Parameters:
      $\{R;S;T;P\}=\{3.0;0.0;3.5;0.5\}$ (region A).
      \label{fig:case5_freq}}
\end{figure}

From the computer simulations shown in \pic{fig:evolution} and
\pics{square}{fig:case5_freq}, we can draw the following conclusions in
agreement with the general discussion in \sect{3} and the discussion of
\pic{fig:evolution} in \sect{2.2}: We observe the formation of
\emph{spatial domains} of cooperators that are separated by narrow
regions of defectors. Only the special case of a singular initial cluster
shown in \pic{square} leads to a regular stationary final pattern where
cooperators and defectors are clearly separated into two domains.  In the
stationary state we find a \emph{stable coexistence} between cooperators
and defectors.  The cooperators become the majority ($f>0.5$) almost
independent of their initial spatial configuration and initial frequency
as shown in \pic{cluster}A and \pic{fig:case5_freq}.  The stationary
frequency is almost constant (with very small periodic changes).

\subsection{Coexistence with a minority of cooperators  - spatial chaos
  (region B)}
\label{4.4}

If only two of the coexistence conditions are satisfied,
i.e. \eqn{invas3} applies, this leads to: 
\begin{equation}
\label{eq:constr2}
\frac{2R+2S}{4} > \frac{T+3P}{4}\;; \quad
\frac{3R}{4}> \frac{2T+2P}{4}\;; \quad
R < \frac{3T+P}{4}
\end{equation}
With the fixed values for $R$ and $S$, we find: 
\begin{equation}
\label{eq:constr22}
T+3P<6 \;; \quad 
2T+2P<9\;; \quad
3T+P>12
\end{equation}
With \eqn{tp}, this eventually results in the solution:
\begin{equation}
\label{eq:soln2}
4-\frac{P}{3}<T<4.5-P \quad \mathrm{if} \quad 0<P<0.75
\end{equation}
This condition defines region \textbf{B} of \pic{phase}.  \Eqn{a-B}
gives the respective payoffs $a_{\theta}^{s}$, \eqn{payoff-both} for the
choosen set of parameters, $\{R;S;T;P\}=\{3.0;0.0;3.9;0.5\}$. The
$a_{\theta}^{s}$ can be used to verify the general considerations of
\sect{3}.
\begin{equation}
  \label{a-B}
  \begin{array}{ccccccccccc}
\quad a_{1}^{4} \quad & \quad a_{1}^{3} \quad & \quad a_{1}^{2} \quad &
\quad a_{1}^{1} \quad & \quad a_{1}^{0} \quad & \; \; & 
\quad a_{0}^{4} \quad & \quad a_{0}^{3} \quad & \quad a_{0}^{2} \quad & 
\quad a_{0}^{1} \quad & \quad a_{0}^{0} \quad \\
3.0 & 2.25 & 1.5 & 0.75 & 0.0 & & 3.9 & 3.05 & 2.2 & 1.35 & 0.5 
  \end{array}
\end{equation}
For the appropriate $\{T,P\}$ values, there is always a
\emph{nonstationary} coexistence between cooperators and defectors where
the cooperators are a \emph{minority}.  This is also shown for the two
different initial conditions, \pic{ini} in the computer simulations of
\pic{fig:ev3_5}, \pic{fig:case3}, \pic{cluster}B and \pic{fig:case3_freq}.
\begin{figure}[htbp]
\hspace{8pt}
\psfig{file=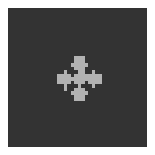,width=2.2cm}(5)\hspace{5pt}
\psfig{file=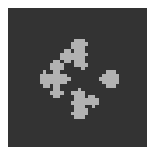,width=2.2cm}(10)\hspace{0.5pt}
\psfig{file=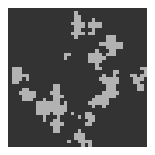,width=2.2cm}(20)\hspace{0.5pt}
\psfig{file=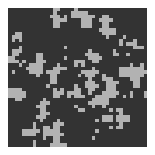,width=2.2cm}(30)\hspace{5pt}
\caption{Time series of the spatial distribution of cooperators (grey)
  and defectors (black) on a CA of size $N=40\times 40$. The time is
  given by the numbers of generations in brackets.  Initial condition:
  One cluster of 9 cooperators, \pic{ini}(left).  Parameters for the
  payoff matrix, \eqn{2payoff}: $\{R;S;T;P\}=\{3.0;0.0;3.9;0.5\}$ (region
  B).  \label{fig:ev3_5}}
\end{figure}
\begin{figure}[htbp]
  \centerline{
  \psfig{file=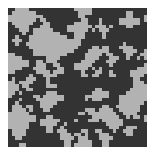,width=2.2cm}\hspace{15pt}
  \psfig{file=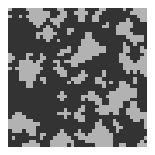,width=2.2cm}\hspace{15pt}
  \psfig{file=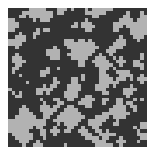,width=2.2cm}\hspace{15pt}
}
\caption{Spatial distribution of cooperators (grey) and
  defectors (black) on a CA of size $N=40\times 40$ after $G=100$
  generations. Note, that there is no stationary pattern asymptotically.
  Initial condition: random spatial distribution of cooperators and
  defectors, left: $f(0)=0.5$, middle: $f(0)=0.75$, right: $f(0)=0.9$.
  Parameters: $\{R;S;T;P\}=\{3.0;0.0;3.9;0.5\}$ (region B).
  \label{fig:case3}}
\end{figure}
\begin{figure}[htbp]
  \centerline{\psfig{file=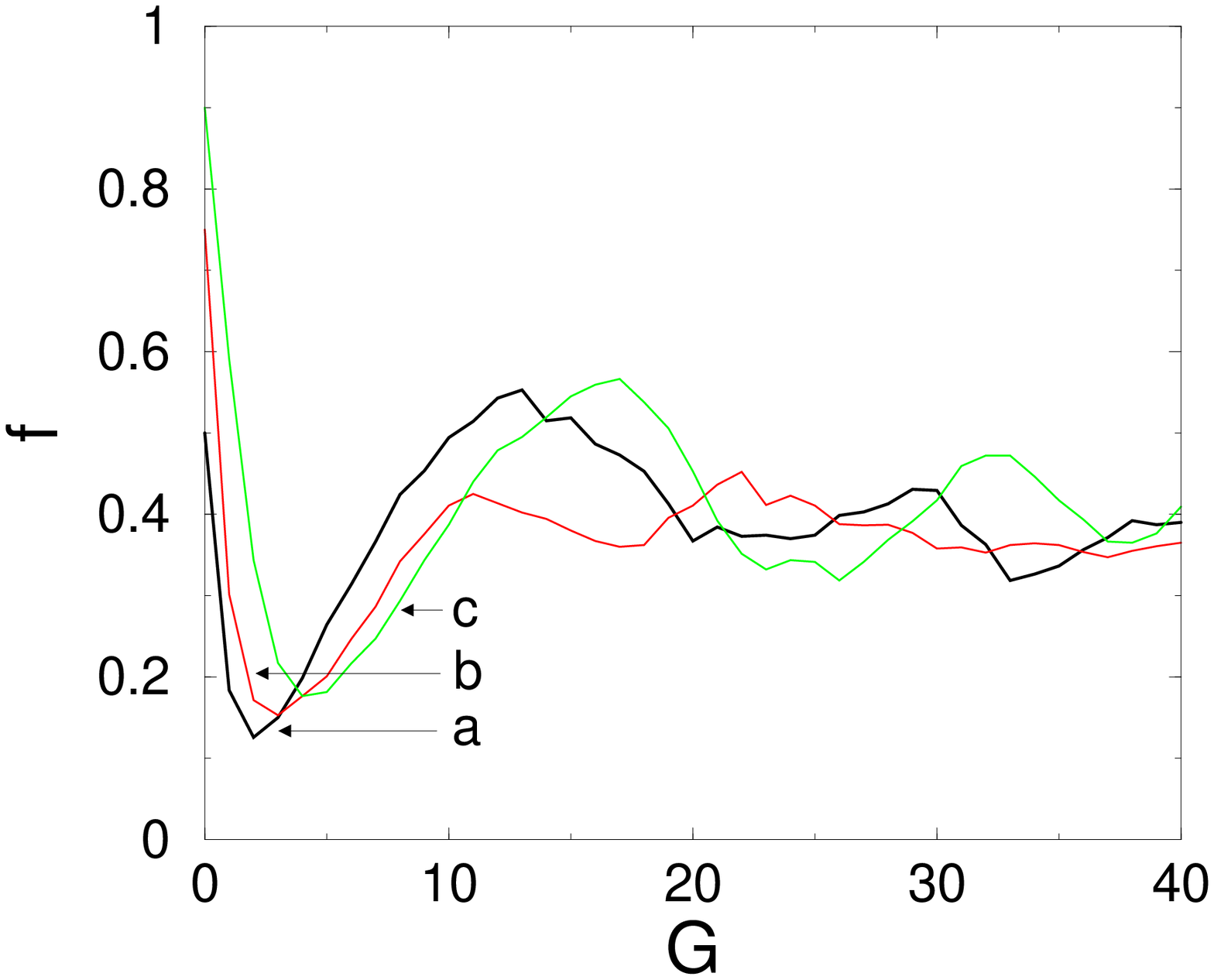,width=7.5cm}}
  \caption{Global frequency of cooperators, $f$, \eqn{nconst}, vs. time
    (number of generations).  The data have been obtained from
    simulations with three different initial frequencies: (a) $f(0)=0.5$,
    (b) $f(0)=0.75$, (c) $f(0)=0.9$. See also \pic{fig:case3} for the
    final spatial distributions. Parameters:
    $\{R;S;T;P\}=\{3.0;0.0;3.9;0.5\}$ (region B).
    \label{fig:case3_freq}}
\end{figure}

The time series of \pic{fig:ev3_5} -- that should be compared to
\pic{square} for parameter region \textbf{A} -- shows that the initial
cluster of cooperators first grows, but then splits up into smaller
clusters, which grow, split up again and may dissappear. This leads to a
non-stationary spatial distribution even in the long run. In fact, it has
been already argued by Nowak \citep{Nowak:93} that this regime can be
characterized as \emph{spatiotemporal chaos}. If we look for the time
dependence of the global frequency in this case, \pic{cluster}B and
\pic{fig:case3_freq}, we find no convergence to a (quasi)stationary value
as in \pic{fig:case5_freq}, but rather large variations over time.
Noteworthy, in the average the global frequency of cooperators is below
0.5, i.e.  they are indeed the minority almost independent of the initial
frequencies. Further, we note that the cooperators still form large
clusters, as shown \pic{fig:case3} that can be compared to
\pic{fig:case5} for region \textbf{A}.

\subsection{Coexistence with cooperators in small clusters (region C)}
\label{4.5}
Eventually, we also show examples for the weaker coexistence condition,
\eqn{invas4}. In this case, the conditions are given as:
\begin{equation}
\label{eq:constr1}
 \frac{2T+2P}{4} > \frac{3R+S}{4} > \frac{T+3P}{4}\; ; \quad
\frac{3T+P}{4} > R> \frac{2T+2P}{4}
\end{equation}
With $R=3$ and $S=0$, this results in:
\begin{equation}
\label{eq:constr12}
T+3P<9 \;; \quad
2T+3P<12\;; \quad
2T+2P>9\;; \quad
3T+P>12 
\end{equation}
where additionally \eqn{tp} applies.  The solution of \eqn{eq:constr12}
is then given by:
\begin{equation}
\label{eq:soln1}
\begin{array}{rclcrcl}
4-P/3 &<T<& 9-3P & \quad \mathrm{if} \quad & 1.5 &<P<& 1.875  \\
4-P/3 &<T<& 6-P & \quad \mathrm{if} \quad & 0.75 &<P<& 1.5  \\
4.5-P& <T<& 6-P & \quad \mathrm{if} \quad & 0.0 &<P<& 0.75 
\end{array}
\end{equation}
These three conditions define region \textbf{C} of \pic{phase}.   \Eqn{a-C}
gives the respective payoffs $a_{\theta}^{s}$, \eqn{payoff-both} for the
choosen set of parameters, $\{R;S;T;P\}=\{3.0;0.0;4.5;1.0\}$. The
$a_{\theta}^{s}$ can be used to verify the general considerations of
\sect{3}.
\begin{equation}
  \label{a-C}
  \begin{array}{ccccccccccc}
\quad a_{1}^{4} \quad & \quad a_{1}^{3} \quad & \quad a_{1}^{2} \quad &
\quad a_{1}^{1} \quad & \quad a_{1}^{0} \quad & \; \; & 
\quad a_{0}^{4} \quad & \quad a_{0}^{3} \quad & \quad a_{0}^{2} \quad & 
\quad a_{0}^{1} \quad & \quad a_{0}^{0} \quad \\
3.0 & 2.25 & 1.5 & 0.75 & 0.0 & & 4.5 & 3.625 & 2.75 & 1.875 & 1.0 
  \end{array}
\end{equation}
For the appropriate $\{T,P\}$ values, the stationary states are
characterized by a \emph{stable} coexistence between cooperators and
defectors where the cooperators survive only in \emph{small clusters},
while the defectors dominate.  This is also shown for the two different
initial conditions of \pic{ini} in the computer simulation of
\pic{fig:ev2_5}, \pic{fig:case1}, \pic{cluster}C and \pic{fig:case1_freq}.
\begin{figure}[htbp]
\centerline{
\psfig{file=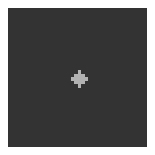,width=2.2cm}(5)\hspace{5pt}
\psfig{file=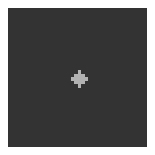,width=2.2cm}(10)\hspace{5pt} {\Large
  $\cdots$}
}
\caption{Time series of the spatial distribution of cooperators (grey)
  and defectors (black) on a CA of size $N=40\times 40$. The time is
  given by the numbers of generations in brackets.  Initial condition:
  One cluster of 9 cooperators, \pic{ini}(left).  Parameters for the
  payoff matrix, \eqn{2payoff}: $\{R;S;T;P\}=\{3.0;0.0;4.5;1.0\}$ (region
  C).  \label{fig:ev2_5}}
\end{figure}
\begin{figure}[htbp]
  \centerline{
  \psfig{file=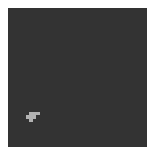,width=2.2cm}\hspace{15pt}
  \psfig{file=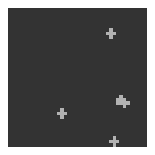,width=2.2cm}\hspace{15pt}
  \psfig{file=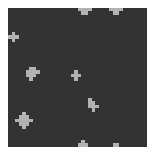,width=2.2cm}\hspace{15pt}
}
\caption{Final (steady state) spatial distribution of cooperators (grey) and
  defectors (black) on a CA of size $N=40\times 40$. Initial condition:
  random spatial distribution of cooperators and defectors, left:
  $f(0)=0.5$, middle: $f(0)=0.75$, right: $f(0)=0.9$. Parameters:
  $\{R;S;T;P\}=\{3.0;0.0;4.5;1.0\}$ (region C).  \label{fig:case1}}
\end{figure}
\begin{figure}[htbp]
  \centerline{\psfig{file=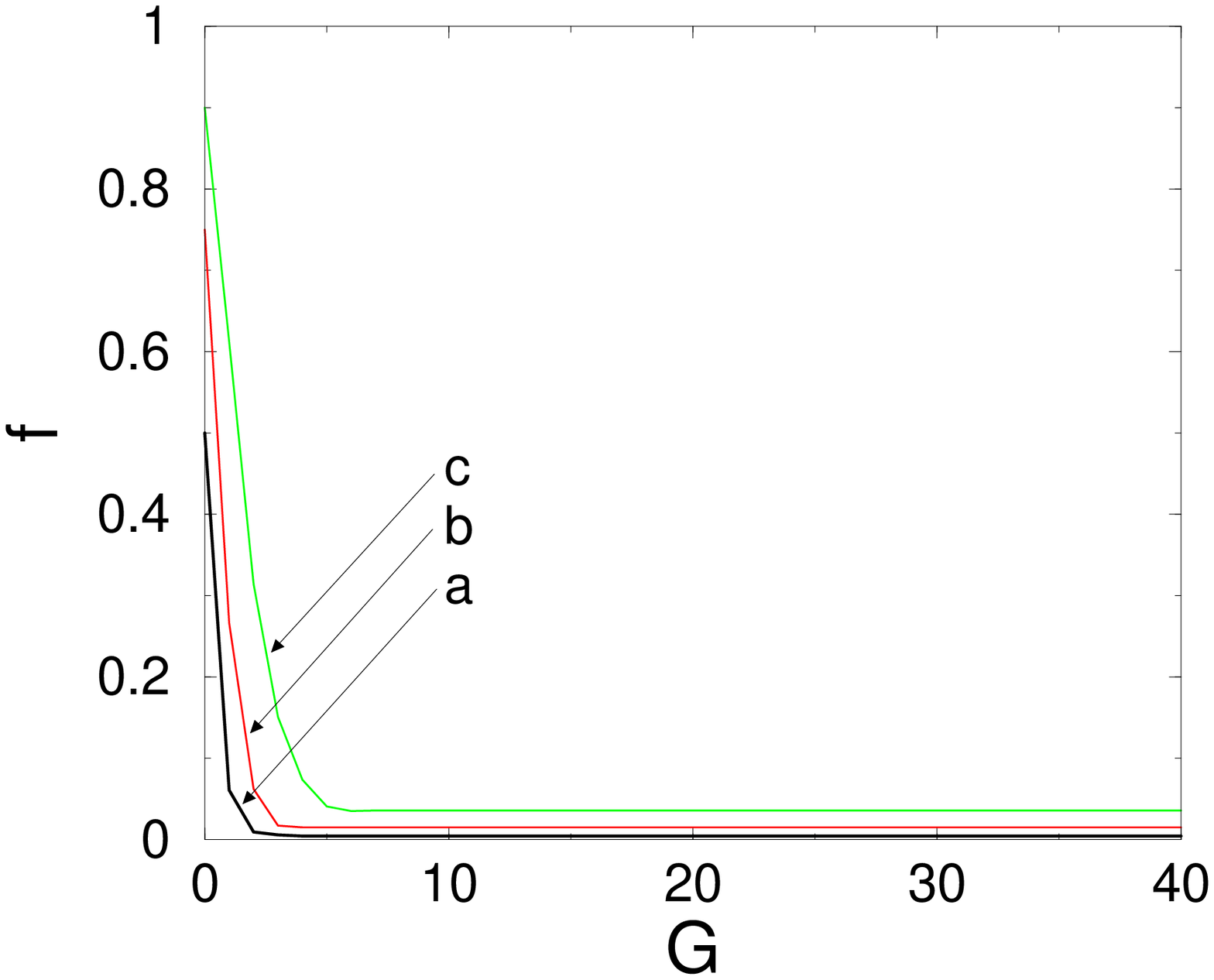,width=7.5cm}}
  \caption{Global frequency of cooperators, $f$, \eqn{nconst}, vs. time
    (number of generations).  The data have been obtained from
    simulations with three different initial frequencies: (a) $f(0)=0.5$,
    (b) $f(0)=0.75$, (c) $f(0)=0.9$. See also \pic{fig:case1} for the
    final spatial distributions. Parameters:
    $\{R;S;T;P\}=\{3.0;0.0;4.5;1.0\}$ (region C).
    \label{fig:case1_freq}}
\end{figure}

\pic{fig:ev2_5} -- that should be compared to \pic{square} for parameter
region \textbf{A} and \pic{fig:ev3_5} for region \textbf{B} -- indicates
that the initial cluster grows only very little in size, i.e.  from 9 to
13 cooperators, and then remains stable at its borders.  However, if the
initial configuration would have more than one cluster, then not all of
them may survive. If two cooperating clusters are very near, then the
configuration $K_0^{4}$ may appear between them, which leads to an
invasion of the defectors in the next step and the further discussion of
\sect{3.2} applies.  Thus, several small clusters may only survive if
there is a certain distance between them. This can be also seen in
\pic{fig:case1} that shows the stationary spatial distributions for
parameter region \textbf{C}, in comparison to \pic{fig:case5} for
region \textbf{A} and \pic{fig:case3} for region \textbf{B}.  We observe
only very small clusters at a certain distance, the number of which
further depend on the initial frequency of cooperators. The evolution of
the global frequency, \pic{cluster}C and \pic{fig:case1_freq}, also
clearly shows that the cooperators survive only as a small minority,
which should be compared to \pic{fig:case5_freq} and
\pic{fig:case3_freq}.

\subsection{Special case of unstable coexistence (region D)}
\label{4.6}
The last case to be discussed is the special case of
\eqn{invas5}-(\ref{invas5c}), which leads to coexistence only for very
special initial conditions, as discussed in \sect{3.2}. From
\eqn{invas5}, we find:
\begin{equation}
\label{eq:inva}
\frac{T+3P}{4} < R < \frac{2T+2P}{4}
\end{equation}
With $R=3$, $S=0$, we have following conditions:
\begin{equation}
\label{eq:inva1}
T<12-3P\;; \quad
T<6-P 
\end{equation}
where additionally \eqn{tp} applies. This leads to the solution:  
\begin{equation}
\label{eq:invsoln}
6-P<T<6 \quad \mathrm{if} \quad 0<P<3
\end{equation}
Similarly, the special case of \eqn{invas5b} with $R=3$, $S=0$ and
\eqn{tp} leads to:
\begin{eqnarray}
  \label{eq:5b1}
\frac{2T+2P}{4}<R \;&;& \quad \frac{3R+S}{4}<\frac{T+3P}{4} \nonumber \\
2T+2P<12 \;&;& \quad 9< T+3P 
\end{eqnarray}
with the solution: 
\begin{equation}
  \label{eq:5b}
3<T<6-P \quad \mathrm{if} \quad 2<P<3
\end{equation}
while the special case of \eqn{invas5c} results in: 
\begin{eqnarray}
  \label{eq:5c}
\frac{3T+P}{4}<R\;&;&\quad \frac{2R+2S}{4}<\frac{T+3P}{4}<\frac{3R+S}{4}
\nonumber \\
3T+P<12 \;&;& \quad  6<T+3P<9 
\end{eqnarray}
with the solution: 
\begin{equation}
\label{d2}
\begin{array}{rclcrcl}
3 &<T<& \frac{\D 12-P}{\D 3} & \quad \mathrm{if} \quad & 1.0 &<P<& 2.0  \\
6-3P& <T<& \frac{\D 12-P}{\D 3} & \quad \mathrm{if} \quad & 0.75 &<P<& 1.0 
\end{array}
\end{equation}
The three solutions (\ref{eq:invsoln}), (\ref{eq:5b}), (\ref{d2})
together define parameter region \textbf{D} of \pic{phase}. Provided the
appropriate $\{T,P\}$ values, we observe in this case only an
\emph{unstable} coexistence of cooperators and defectors, i.e. for an
random initial distribution the dynamics will always lead to an
\emph{invasion} of the defectors until a complete extinction of the
cooperators -- rather similar to the dynamics in the adjacent parameter
region \textbf{E}.  Only for the special case of two domains with a
straight border, the coexistence remains stable.

Looking at the phase diagram of \pic{phase}, we notice that region
\textbf{D} \emph{separates} the regions of stable coexistence
(\textbf{A}-\textbf{C}) from the region of complete invasion of defectors
(\textbf{E}), i.e. \textbf{D} denotes the \emph{transition region}
between the two different stationary regimes, and the border between the
two regions \textbf{C} and \textbf{D} just marks the transition from
stability (i.e. stable coexistence) to instability (i.e. unstable
coexistence).  We further notice that the ``classical'' parameter values
$\{R;S;T;P\}=\{3;0;5;1\}$ are just on the border between regions
\textbf{C} and \textbf{D}, i.e. for a random initial distribution the
classical spatial 5-person game will always lead to the complete invasion
of defectors and the extinction of cooperators, as already observed by
means of computer simulations.

\subsection{Dynamics at the border regions}
\label{4.7}

So far, we have discussed the dynamics for the pure parameter regions,
\textbf{A}-\textbf{E}. However, the question of interest is how the
system behaves if we choose the parameters of the payoff matrix on the
\emph{border} between two such regions. We have already mentioned the
case of the \textbf{C$|$D} border, that marks the transition from (stable)
coexistence to invasion. \pic{f-border} presents results of computer
simulations for the \textbf{A$|$B} and \textbf{B$|$C} border. 
\begin{figure}[htbp]
\centerline{
\psfig{file=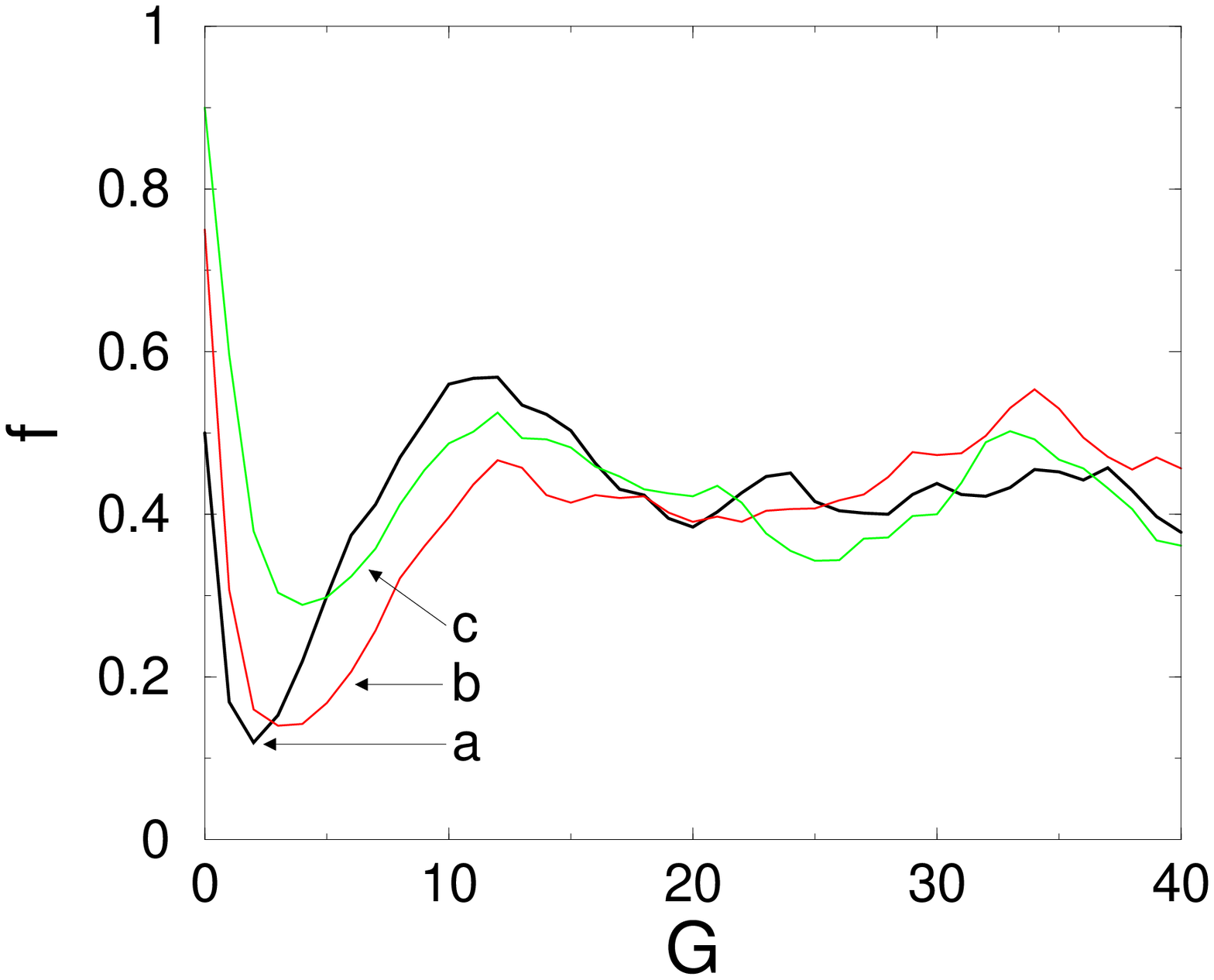,width=6cm}\hfill
\psfig{file=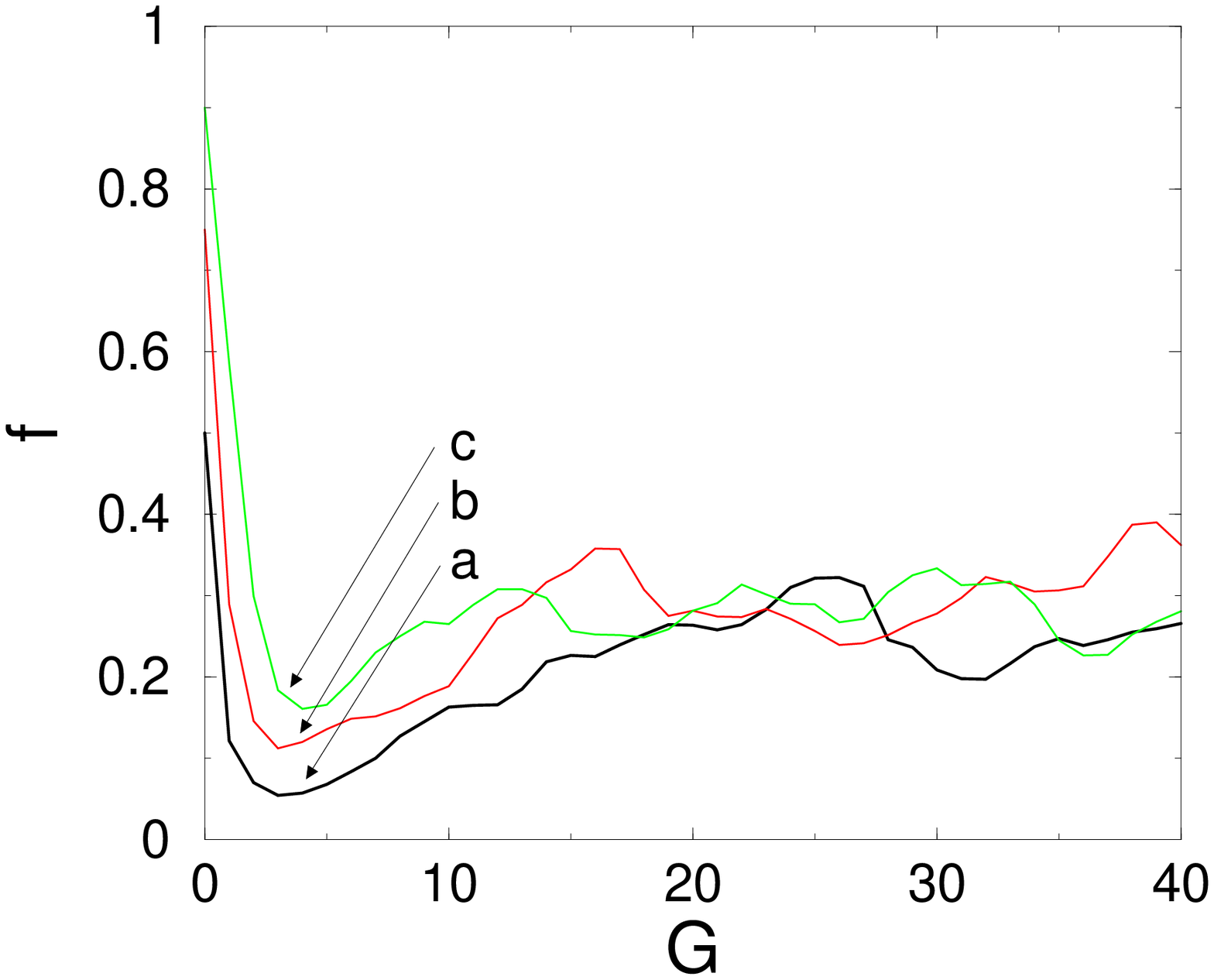,width=6cm}
}
 \caption{Global frequency of cooperators, $f$, \eqn{nconst}, vs. time
   (number of generations).  The data have been obtained from simulations
   with three different initial frequencies: (a) $f(0)=0.5$, (b)
   $f(0)=0.75$, (c) $f(0)=0.9$.  Parameters for $\{R;S;T;P\}$: (left)
   $\{3.0;0.0;3.833;0.5\}$ (border between regions A and B), (right)
   $\{3.0;0.0;4.0;0.5\}$ (border between regions B and C).
    \label{f-border}}
\end{figure}

As we see from the time evolution of the global frequencies, the dynamics
in this case is basically the same as for region \textbf{B}, i.e. there
is no (quasi)stationary regime as for regions \textbf{A} or \textbf{C},
and we may expect a spatiotemporal chaos again.  Despite this, the
average global frequency in the long run is also decreasing when going
from region \textbf{A} to \textbf{C}. In \pic{f-border}(left), i.e. on the
border \textbf{A$|$B} it is below the values of region \textbf{A}, where
the cooperators appear as a majority, \pic{fig:case5_freq}, but above the
values of region \textbf{B}, where they already appear as a minority,
\pic{fig:case3_freq}.  But in \pic{f-border}(right), i.e. on the border
\textbf{B$|$C} it is below the values of region \textbf{B}, but still above
the values of region \textbf{B}, where the cooperators appear as a clear
majority, \pic{fig:case1_freq}.

\subsection{Nowak's results revisited}
\label{4.8}

For the parameter region \textbf{B} we already mentioned the existence of
\emph{spatiotemporal chaos} in the distribution of cooperators and
defectors. This has been first observed by Nowak \citep{Nowak:93} by
means of computer simulations. Based on our detailed theoretical
investigations above, we are now able to derive the critical parameter
values for Nowak's simulations.

Different from the standard values of the prisoner's dilemma,
$\{R;S;T;P\}=\{3;0;5;1\}$, Nowak has used the following values for the
payoff matrix:
\begin{equation}
  \label{nowak}
  \{R;S;T;P\}=\{1;0;b;0\}
\end{equation}
i.e. all payoffs are fixed, while only the payoff for the defecting agent
playing with a cooperating agent is an adjustable value, $b$.  In order
to fulfill the conditions for the payoff, \eqs{pd-ineq1}{pd-ineq2}, $b$
can have only values of $1<b<2$. Further, we notice that the payoff
matrix given by \eqn{nowak} does not \emph{strictly} fulfill the
conditions of the prisoner's dilemma, because of $P=S$ in this case.

We can now apply the conditions derived for the different regions to the
payoff matrix of \eqn{nowak}. The results are given in Table \ref{b-nowak}.
\begin{table}[htbp]
    \caption{Parameter range of the payoff $b$, \eqn{nowak}
to obtain the different spatial patterns in the asymtotic state.}
\begin{center}
   {
         \begin{tabular}[c]{llr@{$\;$}c@{$\;$}l@{$\;$}}
\hline \hline
stationary state & condition & \multicolumn{3}{c}{range of $b$ values} 
\\ \hline
\textbf{A}: coexistence with large domains of cooperators & 
\eqn{eq:constr3} &
1.0 &$<b<$ & 1.33 \\
\textbf{B}: coexistence with spatial chaos & \eqn{eq:constr2} & 
1.33 &$< b <$& 1.5 \\
\textbf{C}: coexistence with small clusters of cooperators &
\eqn{eq:constr1} & 1.5 &$<b<$ & 2.0 \\
\hline \hline
    \end{tabular} 
}   
\end{center}
   \label{b-nowak}
\end{table}

We note again that these regions have been found in \citep{Nowak:93} by
means of computer simulations, while we have confirmed the results based
on an analytical investigation of the stability conditions. Further, we
want to point out that the other two dynamical regimes, i.e. \textbf{E}:
complete invasion of defectors and \textbf{D}: unstable coexistence, do
not exist for the payoff matrix, \eqn{nowak}.

\section{Interactions in a larger neighborhood}
\label{5}

In this section, we want to implement the ``exact'' (or complete)
5-person game for the one-shot PD game on the square lattice - which is
to our knowledge not investigated so far.  Since the payoff matrix for
the 5-person game has been computed in this paper using the concept of
the 2-person game, \sect{2.2}, we will continue to explain the dynamics
of the 5-person game in terms of 2-person games.  So far, we have
investigated the spatial 5-person game under the assumptions given in
\sect{2.2}, namely (i) decomposition into independent, simultaneous
2-person games, (ii) each agent only plays with its $m$ nearest
neighbors. This has reduced the game to $m$ 2-person games played by each
agent.

In the following, we want to discuss a different variant of the game
which instead of (ii) assumes that in a neighborhood of $n=5$ \emph{each
  agent} plays a 2-person game with \emph{every other agent} in this
neighborhood.  For a given neighborhood, this increases the number of
2-person games to $m\times n/2=10$. We further have to take into account
that agent $i$ itself is additionally part of $m$ different neighborhoods
of size $n=5$ centered around its nearest neighbors at the positions
$i_{j}$ $(j=1,...,m)$. This results in the consideration of a larger
neighborhood of size $n=13$ that includes also the $r$ second nearest
neighbors of agent $i$ (see \pic{lattice}). The total number of 2-person
games played independently in this larger neighborhood is given by
$m\,n^{2}/2=50$, but we can easily verify that agent $i$ at position
$j=0$ participates only in 20 of these.  Specifically, he plays 4 games
in his ``own neighborhood'' $(j=0)$ and 4 times 4 games in the
neighborhoods of his neighbors $(j=1,...,m)$.  In these 20 games, he
meets always \emph{twice} (in two different 5-person games) both with his
$m$ nearest neighbors and with the neighboring agents at the positions
$j=6,8,10,12$, but only once with the agents at the second nearest
neighbor positions $j=5,7,9,11$ (cf.  \pic{lattice}).

This in turn results in a different payoff, dependent on the number of
cooperators and defectors in the larger neighborhood of agent $i$. Since
the payoff is again calculated from the independent 2-person games, we
can still use \eqn{payoff-i} for $a_{i}$, but we have to recalculate the
fractions $z_{i}^{1}$ and $z^{0}_{i}$ according to the modified game.
I.e. \eqn{sum} has to be replaced by:
\begin{equation}
  \label{sum2}
z_{i}^{\theta} = \frac{1}{(m+r)}\left(2\sum_{j=1}^{m+r}
\delta_{\theta\theta_{i_{j}}} - \sum_{j\in5,7,9,11}
\delta_{\theta\theta_{i_{j}}} \right)
;\;
z_{i}^{(1-\theta)} = 1 - z_{i}^{\theta} \;\; (\theta\in \{0,1\} )
\end{equation}
\Eqn{sum2} can be interpreted in a way that agent $i$ -- in addition to
his nearest neighbors -- now also interacts with his second nearest
neighbors, but if they are not adjacent to his place (i.e.
$j\in5,7,9,11$), this interaction occurs less frequently -- or, the
payoff is counted with a smaller weight, respectively.

We do not intend to give here a detailed analysis of this modified case,
the effect of the weighted interaction in the larger neighborhood shall
be rather demonstrated by some computer simulations that allow a
comparison to the previous results.  For the dynamics of the modified
game the update rule, \eqs{update0}{update} still applies, i.e. agent $i$
adopts the $C$ or $D$ behavior from the agent that received the highest
payoff in its neighborhood of $n=5$.  The parameters of the payoff
matrix, \eqn{2payoff} have been choosen in region \textbf{A} of the phase
diagram, \pic{phase}, where we found the spatial coexistence defectors
(as the minority) and cooperators (as the majority), the payoff values
$a_{\theta}^{s}$ are given by \eqn{a-A}.

For the computer simulations, we have again used the initial spatial
configurations shown in \pic{ini}, i.e. either one small cluster of
cooperators, or a random intial distribution of cooperators and defectors
with different intial frequencies. \pic{fig:evolution_5} -- that shall be
compared to \pic{square} -- shows that in the modified case the domain of
cooperators considerably increases, i.e. it reaches $f=0.95$ compared to
$f=0.52$.  Further the dynamics occurs faster (compare the snapshot after
30 generations in \pic{square} with the one after 20 generations in
\pic{fig:evolution_5}) and the border between the two domains remains
planar during the evolution.
\begin{figure}[htbp]
\hspace{8pt}
\psfig{file=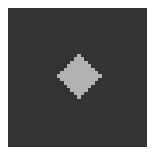,width=2.2cm}(5)\hspace{5pt}
\psfig{file=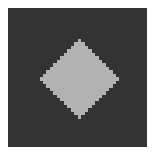,width=2.2cm}(10)\hspace{0.5pt}
\psfig{file=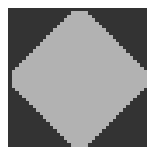,width=2.2cm}(20)\hspace{0.5pt}
\psfig{file=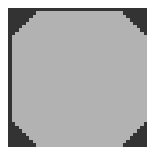,width=2.2cm}(30)
\caption{Time series of the spatial distribution of cooperators (grey)
  and defectors (black) for the modified game, \eqn{sum2}. The time is
  given by the numbers of generations in brackets.  Initial condition:
  One cluster of 9 cooperators, \pic{ini}(left).  Parameters for the
  payoff matrix, \eqn{2payoff}: $\{R;S;T;P\}=\{3.0;0.0;3.5;0.5\}$ (region
  A).
  \label{fig:evolution_5}}
\end{figure}
\begin{figure}[htbp]
  \hspace{8pt} \psfig{file=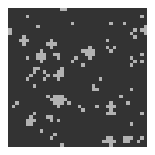,width=2.2cm}(1)\hspace{5pt}
  \psfig{file=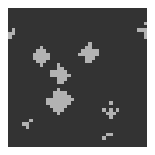,width=2.2cm}(3)\hspace{5pt}
  \psfig{file=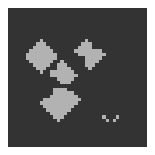,width=2.2cm}(5)\hspace{5pt}
  \psfig{file=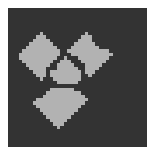,width=2.2cm}(7)\hspace{5pt}
\caption{Time series of the spatial distribution of cooperators (grey)
  and defectors (black) for the modified game, \eqn{sum2}. The time is
  given by the numbers of generations in brackets.  Initial condition:
  $f(0)=0.5$, random spatial distribution of cooperators and defectors,
  \pic{ini}(right).  Parameters: $\{R;S;T;P\}=\{3.0;0.0;3.5;0.5\}$
  (region A). We note that the further evolution leads to a final spatial
  distribution similar to the one shown in \pic{fig:ess_case12}(left).
\label{1-7}}
\end{figure}
\begin{figure}[htbp]
  \centerline{\psfig{file=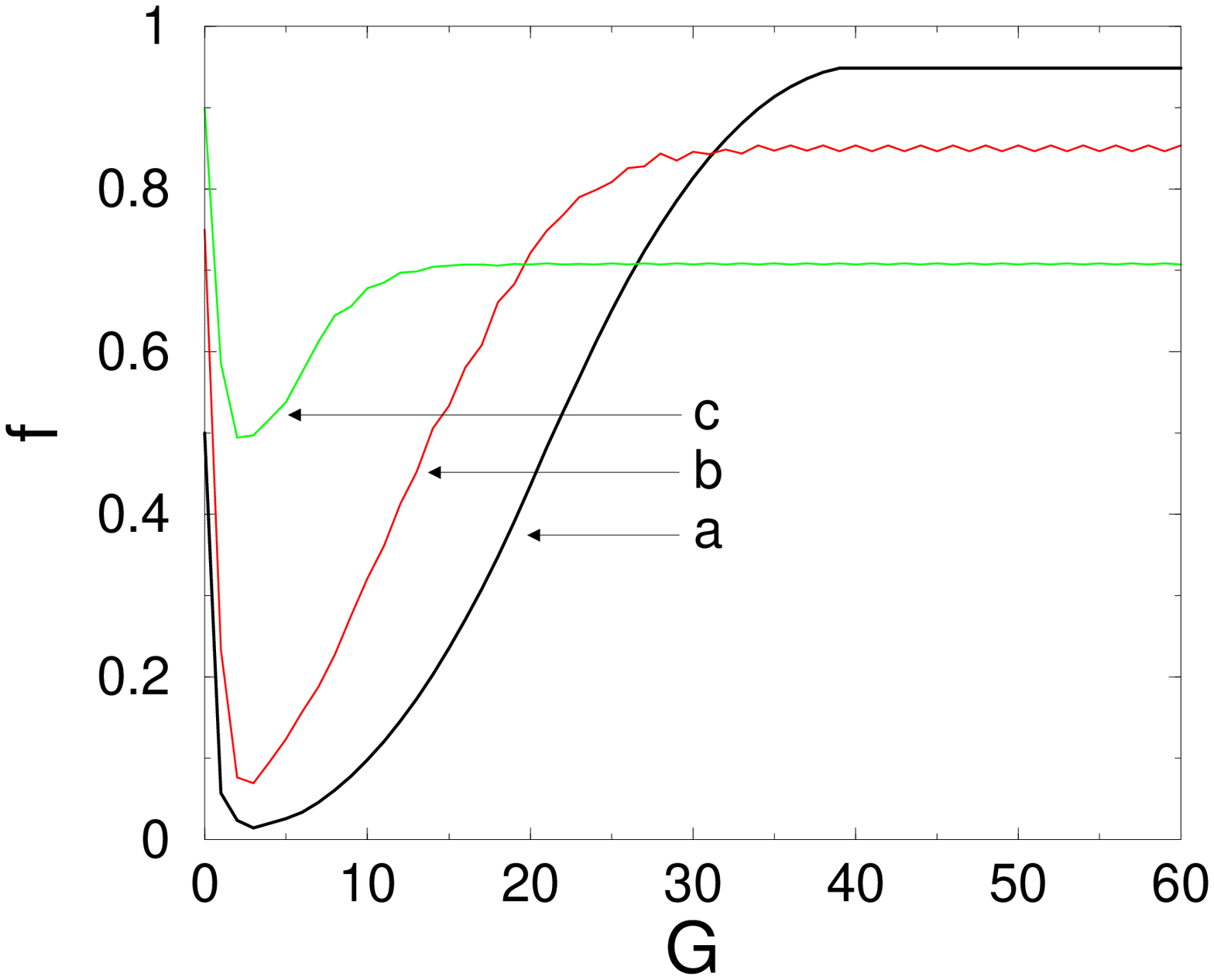,width=7.5cm}}
    \caption{Global frequency of cooperators, $f$,
      \eqn{nconst}, vs. time (number of generations) for the modified
      game, \eqn{sum2}.  The data have been obtained from simulations
      with three different initial frequencies: (a) $f(0)=0.5$, (b)
      $f(0)=0.75$, (c) $f(0)=0.9$. See also \pic{fig:ess_case12} for the
      final spatial distributions. Parameters:
      $\{R;S;T;P\}=\{3.0;0.0;3.5;0.5\}$ (region A).
      \label{fig:12neigh_freq}}
\end{figure}
\begin{figure}[htbp]
  \centerline{
  \psfig{file=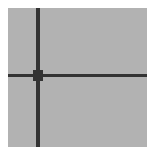,width=2.2cm}\hspace{15pt}
  \psfig{file=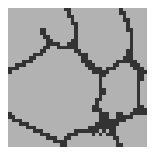,width=2.2cm}\hspace{15pt}
  \psfig{file=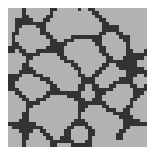,width=2.2cm}\hspace{15pt}
}
\caption{Final (steady state) spatial distribution of cooperators (grey) and
  defectors (black) for the modified game, \eqn{sum2}. Initial condition:
  random spatial distribution of cooperators and defectors, left:
  $f(0)=0.5$, middle: $f(0)=0.75$, right: $f(0)=0.9$. Parameters:
  $\{R;S;T;P\}=\{3.0;0.0;3.5;0.5\}$ (region A).  \label{fig:ess_case12}}
\end{figure}

The influence of the larger neighborhood on the evolution of the spatial
patterns becomes more visible when we start the simulations from a random
initial distribution, as shown in \pic{1-7}. The corresponding evolution
of the global frequency $f(G)$ is shown in \pic{fig:12neigh_freq}(a).
These figures should be compared to \pic{fig:evolution} and
\pic{fig:freq}. Starting with an initial frequency $f(0)=0.5$, we find
that the cooperators during the first three generations almost cease to
exist, they survive only in a few rather small clusters. But it is worth
to be noticed that this situation then changes drastically: the small
clusters grow into a few large domains that are separated by only tiny
borders formed by defectors. Thus, compared to the previous simulations,
we now find (i) less separated domains, and (ii) a much greater
domination of the cooperators in the final state, i.e. $f=0.95$ in this
case. Interestingly, this domination becomes the greater, the less the
inital frequency is (cf. \pic{fig:12neigh_freq}) -- but a the same time
also the risk increases that cooperators die out during the first
generations. \emph{If} they survive, their ``comeback'' is overwhelming.

\pic{fig:ess_case12} shows the spatial final distributions that
correspond to the different frequencies in \pic{fig:12neigh_freq}. We
notice that the different initial frequencies $f(0)$ do not only have a
strong impact on final frequencies, but also on the number of domains
formed during the evolution. The higher the \emph{initial} frequency of
cooperators, the less is the \emph{final} frequency of cooperators and
the more they are splitted into \emph{separated domains}. This effect was
not so pronounced for the simulations shown in \pic{fig:case5} -- thus,
we may conclude that the modified game, i.e. the consideration of the
spatially heterogeneous interaction in the larger neighborhood may lead
to nontrivial diversification in the spatial dynamics, which will be
investigated in a forthcoming paper.

\section{Conclusion}

In this paper we have investigated the spatial organization of
cooperating and defecting agents distributed on a square lattice.  From
his interaction with other agents, each agent receives a payoff that can
be compared to the payoffs of the other agents. Based on this outcome
the agent can in the next generation adopt the more successful behaviour,
either to cooperate (C) or to defect (D).

Different from a mean-field approach that assumes a panmictic population
where each agent interacts with every other agent, we have considered a
spatially restricted case, where each agent interacts only with his
neighbors. This results in a \emph{spatial 5-person game} that has been
discussed in two variants: (i) each agent interacts only with his four
nearest neighbors, (ii) each agent in the neighborhood of 5 interacts
with any other agent in this neighborhood. Such a distinction leads to a
different number of 2-person games played simultaneously, but
independently by each agent in a given neighborhood, i.e.  4 games in the
first case and 10 games in the second case.

The main part of the paper is devoted to the first case. Based on the
exact calculation of the payoffs for the possible encounters in a given
neighborhood, we have derived analytical expressions (in terms of
inequalities) to characterize the different spatial organizations of
cooperators and defectors. The results are concluded in a \emph{phase
  diagram} for the $\{T,P\}$ parameter space of the possible payoffs for
defectors, \pic{phase}. We could identify \emph{five} different dynamical
regimes (A-E), each characterized by a distinct spatiotemporal dynamics
and a corresponding final spatial distribution. We found that for
arbitrary initial conditions parameters choosen from regions E and D will
always lead to a spatial \emph{invasion} of the defectors and eventually
results in a complete \emph{extinction} of the cooperators.  This steady
state agrees with the steady state of the  mean-field dynamics, where the
invasion of defectors is the only possible outcome. 
I.e., if the values of the payoff matrix are choosen from regions
D and E,  then space plays no longer a role in the determination of the
steady state, that is the same regardless of whether the interaction
occurs only between nearest neighbors or between all agents in the
system.

In addition to that we could also identify parameter regions (A, B, C)
characterized by a \emph{spatial coexistence} between cooperators and
defectors. A detailed analysis could reveal the conditions under which
the cooperators could survive either as a \emph{majority} organized in
large spatial domains, or as a \emph{minority} organized in small
non-stationary domains or in small clusters.

The analytical results obtained have been further applied to a spatial
game introduced by \citet{Nowak:93}. The parameter findings obtained
there by means of computer simulations could be confirmed by our
analytical approach.

In the last part of the paper, we have focussed on the variant (ii) of
the 5-person game, where each agent interacts with any other agent in his
neighborhood. If this ``true'' 5-person game is put on a rectangular
lattice -- as we did here to our knowledge for the first time -- it
results in interesting effects. Since each agent is part of different
spatial 5-person games (played in his immediate neighborhood) the
modification eventually leads to the consideration of the second-nearest
neighbors.  We could show that the interaction in this larger
neighborhood can be described as a \emph{non-uniform} spatial game, where
each agent plays 2-person games more frequently with his adjacent
neighbors than with his second-nearest neighbors. As the result, we found
by means of computer simulations that during the first stages of the
evolution the risk considerably increases that the cooperators ceases to
exist, but if they survive, they can become a much larger majority than
in the ``simplified'' case (i). In turn, it is the defectors that only
survive in thin borders.  This is very interesting since true 5-person
games are rather complex, and it is known from mean-field investigations
that achieving cooperation becomes even more difficult in multi-person
games. Thus, the analytical investigations in this paper shall be also
applied to the true spatial 5-person game, in order to reveal its
critical conditions.

We can conclude that space indeed plays a definite role in the evolution
of cooperation, because a spatially restricted interaction may lead to a
global cooperation, even if individual rationality tilts toward
defection. Even in a one-shot PD game it is possible to find a majority
of cooperators, provided a locally restricted interaction is considered.

\bibliography{web-acs}
\end{document}